\documentclass[]{JFM-FLM_Au}

\usepackage{tabularx}
\usepackage{graphicx}
\usepackage{subcaption}
\usepackage{enumitem}
\usepackage{booktabs}
\usepackage{float}

\newcommand{\bm}{\mathbf}
\newcommand{\figpath}{.}
\newcommand{\ii}{\mathrm{i}}

\lefttitle{L. C. Toledo, A. Gesla and E. Yim}
\righttitle{Journal of Fluid Mechanics}

\title{Linear stability and structural sensitivity of a swirling jet in a Francis turbine draft tube}

\author{Lester C. Toledo\aff{1}, Artur Gesla\aff{1} \and Eunok Yim\aff{1}}

\affiliation{\aff{1}Hydro Energy and Applied fluid Dynamics (HEAD) Laboratory, École Polytechnique Fédérale de Lausanne (EPFL), 1015 Lausanne, Switzerland}

\corresau{E. Yim, \email{eunok.yim@epfl.ch}}
\begin{document}
\maketitle

\begin{abstract}
	Motivated by the need to better understand flow unsteadiness in hydraulic turbines, we perform a local linear stability and adjoint-based sensitivity analysis of the turbulent swirling jet at the outlet of a Francis turbine. We use  measured mean flow and turbulence profiles at several operating conditions (below, at, and above the best efficiency point (BEP) flow rate) and perform a stability analysis.
	Incorporating eddy viscosity $\nu_t$ into the analysis strongly damps inviscid growth rates and restricts instability to low azimuthal modes $m\in [-1,2]$, in better agreement with experiments. Three turbulent viscosity closures (constant, mixing-length and measured $k - \varepsilon$ based) yield similar spectra, with close agreement between mixing length and measured models, all identify partial load (0.92 BEP) as the most unstable regime. 
	Sensitivity results show that axial velocity modifications primarily control growth rates, whereas azimuthal velocity changes mainly shift frequencies. We also derive the sensitivity kernel of the spectrum to turbulent viscosity modifications and find that  spatial variations of eddy viscosity are essential for predicting the unstable mode range. The predictions accurately estimate stability changes for small variations in operating point. 
	We further analyze the flow using classical inviscid swirling jet instability criteria (the generalized Rayleigh discriminant) and WKB analysis to predict the stability to broader operating points and reconcile these results to the stability and sensitivity analyses. 
	The approach used in this study is fast and simple to model, but it neglects draft tube geometry (non-parallel effects), motivating future global stability and sensitivity analyses.
\end{abstract}
	
	
	\section{Introduction}
	\label{sec:intro}
\begin{figure}
	\centering
	\includegraphics[width=\textwidth]{\figpath/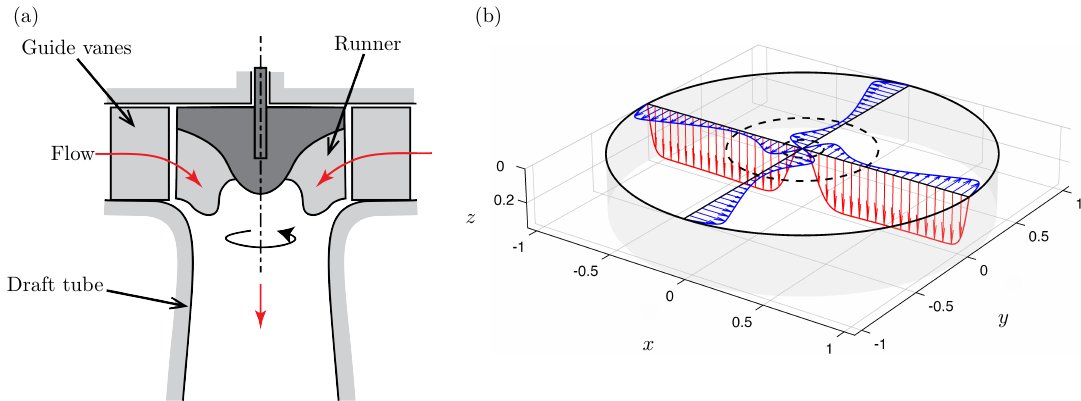}
	\caption{\label{fig:baseflow_idx2} 
		(a) Sketch of a Francis turbine. The draft tube is located downstream of the turbine runner. (b) Three-dimensional visualization of the velocity distribution at the inlet of the draft tube mean swirling flow profile for $0.92$ BEP. The azimuthal $U_\theta$ and axial $U_z$ velocity distributions are shown in blue and red arrows, respectively. The vortex core radii ($R_1$ and $R_2$) associated with the primary swirling regions, are shown as dashed circles.}
\end{figure}

	The increasing integration of renewable energy sources into power systems has required hydraulic turbines to operate in a wider range of conditions to maintain grid stability. The Francis turbine, which is the most widely used type in hydropower plants, experiences performance degradation when operated away from its best efficiency point (BEP) primarily due to flow instabilities \citep{khullar2025impact}. At partial flow rate (part-load, lower than BEP flow rate), a strong residual swirl develops at the draft tube inlet (downstream of the turbine, shown in figure \ref{fig:baseflow_idx2}a) due to the imbalance between the swirl imparted by the guide vanes and the angular momentum extracted by the turbine runner \citep{escudier1987confined}. This condition leads to the formation of a spiral rotating vortex rope, originating from the roll-up of a strong shear layer between the stagnant central region of the draft tube and the surrounding highly swirling flow \citep{seifi2024linear}. As the flow expands within the conical section of the draft tube, the swirling motion decelerates, leading to vortex breakdown. This phenomenon is recognized as the primary source of strong flow instabilities that cause pressure fluctuations and power oscillations during part-load operation \citep{arpe2009experimental}. Extensive research efforts have  been conducted to understand, model, and mitigate this unsteady flow phenomenon in hydraulic turbines.
	
	\cite{rheingans1940power} and \cite{deriaz1960contribution} were the first to experimentally observe the helical, precessing vortex rope under part-load operating conditions. \cite{nishi1980study} later further characterized this phenomenon, showing that, with increasing inlet flow swirl, a vortex core develops and evolves into this spiral structure rotating around the draft tube axis. At very low flow rate around 0.70-0.85 BEP where pressure fluctuations occur at higher frequencies, the vortex rope exhibits an elliptical cross section and a self-rotating motion \citep{nicolet2011experimental}. Further experiments were performed using different experimental techniques such as by measuring unsteady wall pressure \citep{arpe2002pressure}, by laser Doppler velocimetry (LDV), and by two-phase flow particle image velocimetry (PIV) \citep{ciocan2007experimental, iliescu2008analysis, arpe2009experimental, favrel2015study, muller2016measurement}. Simplified models of the instantaneous vortex rope have also been analytically formulated to more effectively capture and describe this phenomenon \citep{dorfler1980modele, fanelli1989vortex, kuibin2010validation}. 
		
	\cite{Muri02} explained that the draft tube efficiency drop under off-design operations is due to a global instability directly triggered by flow rate variation (see \cite{mauri2004werle}). \cite{Susan06} further investigated this efficiency loss by examining the associated sudden decrease in the pressure recovery coefficient near the BEP flow rate and extended the analysis across a wide range of flow rate. They demonstrated that the mean complex swirling flow in the Francis turbine draft tube can be represented by a superposition of three distinct vortices  \citep{Susan10}. Using these fitted velocity profiles, they performed a linear spatial stability analysis of the axisymmetric mode ($m=0$) which revealed that the flow becomes sensitive to axisymmetric disturbances whenever abrupt changes in draft tube pressure recovery occur.
	
	Several studies have applied hydrodynamic stability analysis to swirling flow profiles. \cite{wang1996columnar, wang1996noncolumnar}  studied the linear stability of an inviscid, axisymmetric, and rotating columnar and non-columnar flows in a finite length pipe, revealing the relation between vortex flow stability and axisymmetric vortex breakdown phenomenon. \cite{delbende1998absolute} investigated the absolute/convective instability properties of Bachelor vortex. \cite{gallaire2003instability, gallaire2003mode} studied the mode selection mechanism of a combined vortex profiles described in \cite{carton1989barotropic} (azimuthal velocity) and \cite{monkewitz1988note} (axial velocity). Further studies on swirling flow instability extend from idealized swirling wakes in pipes \citep{wang1997dynamics, meliga2011control} to the nonlinear evolution of a viscous swirling jet \citep{delbende2005nonlinear}. 
\cite{Oberleithner2011} experimentally observed a turbulent swirling jet undergoing vortex breakdown, exhibiting a self-excited global oscillatory mode arising from a supercritical Hopf bifurcation. \cite{Muller_2020} later applied linear global and adjoint analyses to identify the most receptive region of the flow and implemented open-loop control to mitigate the instability on a highly turbulent swirling jet.
	In the context of a Francis turbine draft tube, linear stability tools have been explored to explain the formation of the rotating vortex rope and the associated vortex breakdown. \cite{zhang2005physical} performed a local stability analysis by fitting their computed velocity profiles to a q-vortex model. \cite{topor2012localization} conducted a numerical simulation based on orthogonal decomposition method coupled with an absolute and convective analysis, yielding a good prediction of the axial wavelength and frequency of the vortex rope. \cite{pochyly2009assessment} carried out a global inviscid stability analysis on the Reynolds-averaged Navier-Stokes solution and found multiple unstable modes including the vortex rope frequency.
	
	Linear stability analyses neglecting the contribution of turbulent Reynolds stresses often lead to ambiguous identification of the most dominant unstable mode in turbulent flows \citep{Viola14}. To close the linearized equations for the velocity field, early studies have employed eddy viscosity models to take into account the turbulent nature of the flow \citep{reynolds1972mechanics, bottaro2006formation, crouch2007predicting, cossu2009optimal, meliga2012sensitivity}. As a first step in including the turbulent effect of the flow in linear stability analysis, the so-called frozen eddy viscosity approach, where eddy viscosity is prescribed as part of the baseflow but not subjected to perturbation, has been implemented successfully \citep{cossu2009optimal, hwang2010amplification, cossu2022onset, kashyap2024linear}. \cite{Viola14} investigated the hub vortex instability in a wind-turbine wake using both a uniform eddy viscosity model and mixing-length models to account for turbulence in the stability analysis. Their results showed a significant reduction in the number of unstable modes when a turbulent diffusion model was introduced, and the instability frequency was also well predicted.  \cite{pasche2017part} adopted this approach to the turbulent mean flow in a draft tube by including a turbulent viscosity model in their linear stability analysis. Their findings successfully identified the vortex rope as a globally unstable eigenmode and reproduced its frequency and spatial structure, closely resembling the spiral vortex breakdown observed in experiments. \cite{Muller_2021} and \cite{Litvinov_2022} predicted the onset of the precessing vortex core in the draft tube of Francis turbines using both global and local stability analysis approaches. These studies indicate that accounting for turbulence via an effective (eddy) viscosity is important for correctly capturing the vortex rope instability in turbulent swirling jet.
	 However, the sensitivity of the turbulent mean flow in the draft tube to turbulent viscosity variations remains unexplored, despite its relevance for the operational flexibility of Francis turbines.
	
	In this study, we perform a local linear stability analysis of the experimentally measured turbulent mean flow at the outlet of the Francis turbine by \cite{Susan06} using three eddy viscosity closures: uniform eddy viscosity, a mixing-length formulation and a $k-\varepsilon$ based model. In addition, we conduct a Wentzel-Kramers-Brillouin (WKB) analysis to assess how well asymptotic results developed for swirling jet profiles capture the instability characteristics of the present baseflow \citep{Billant_Gallaire_2013} and to quantify the influence of viscosity on these predictions.
	To further characterize the various turbine flow, sensitivity analyses to baseflow and turbulent viscosity modifications are performed. Following \cite{Marquet08}, we first investigate the sensitivity of the unstable mode to baseflow modifications, which is relevant when the operating point changes and the mean flow is altered. We then derive a formulation for the sensitivity to turbulent viscosity variations, in order to quantify the influence of the eddy viscosity on the predicted stability characteristics. The sensitivity analyses are then performed for the entire axial wavenumber space, allowing growth rate and frequency curve predictions for a modified baseflow. A supplementary analysis with a reduced turbulent length scale is undertaken to further clarify how modifications of the turbulent viscosity reshape the stability characteristics of the flow. Finally, the evolution of the baseflow and the corresponding stability characteristics across the flow rate spectrum is analyzed by comparing the predictions of early inviscid theories and the current linear stability analysis.
		
	The paper is organized as follows. The problem formulation and theoretical framework are presented in \S \ref{sec:problem}, including the linear stability analysis, WKB analysis and the sensitivity analysis with respect to baseflow modifications. This section also derives the sensitivity to turbulent viscosity modifications for a general, unspecified flow before providing its expanded form for the considered swirling flow profile. The numerical methodology based on the Finite Element Method (FEM) is then described in \S \ref{sec:numerical}. The results of the linear stability, WKB, and sensitivity analyses, including the sensitivity distribution across the entire axial wavenumber space, are discussed in \S \ref{sec:results}. This section also presents the prediction of growth rate and frequency curves of a modified baseflow using the present sensitivity analysis results, as well as an additional study examining the effect of a reduced turbulent length scale on the stability characteristics. The stability of the swirling baseflow is further examined across the entire flow rate spectrum.
Finally, the conclusions are laid out in \S \ref{sec:conclusion}.

\section{Problem formulation}
\label{sec:problem}
\subsection{Baseflow profile}
\label{sec:baseflow}
We use the  mean velocity profile as the baseflow measured at the outlet of a Francis turbine runner (upstream of the draft tube) by \cite{Susan06, Susan10}.
Figure \ref{fig:baseflow_idx2}(a) shows a schematic sketch of a Francis turbine showing the location of the draft tube. Figure \ref{fig:baseflow_idx2}(b) provides a three-dimensional visualization of the velocity distribution of the baseflow profile for 0.92 BEP flow rate. It shows the radial distribution of the swirling flow (blue) and the jet axial flow (red). \cite{Susan06} proposed a mathematical fitting of the mean swirling flow into three distinct vortices, namely: a rigid body motion and two Batchelor vortices with two different core radii. Radial, azimuthal, and axial components of the baseflow $\mathbf{U}_b=[U_r, U_\theta(r), U_z(r)]$ read:  
\begin{eqnarray}
	U_{r} & = & 0, \label{eq:vr}\\
	U_{\theta} & = & \Omega_0 r+ \frac{\Omega_1 R_1^2}{r} \left[1- \exp\left( \frac{-r^{2}}{R_1^2} \right) \right] + \frac{\Omega_2 R_2^2}{r} \left[1- \exp\left( \frac{-r^{2}}{R_2^2} \right) \right], \label{eq:vt}\\
	U_{z} & = & W_0+W_1 \exp \left( \frac{-r^{2}}{R_1^2} \right) + W_2 \exp \left( \frac{-r^{2}}{R_2^2} \right),\label{eq:vz}
\end{eqnarray} 
where the swirl parameters $(\Omega_{0,1,2}, R_{1,2}, W_{0,1,2})$ are given for the different turbine operating points in Table \ref{tab:baseflow_params} (extracted from \cite{Susan06, Susan10}). For all of the operating points, $U_r$ is zero. For $U_\theta$, the second and third terms are the counter-rotating ($\Omega_1<0$) and co-rotating vortices ($\Omega_2>0$), respectively, with respect to the solid body rotation (first term). It can also be expressed as $U_\theta = \Omega(r) r$, where $\Omega(r)$ is the angular velocity. For $U_z$, the second and third terms are the co-flowing ($W_1>0$) and counter-flowing ($W_2<0$) jets, respectively, with respect to the axial freestream velocity ($W_0$ is positive for downward direction). The radius $r$ and velocity $\mathbf{U}$ are non-dimensionalized using the turbine outlet radius $R_0 = 20$ cm and the turbine velocity at this radius ($V_{0} = 1000$ rpm $\times 2\pi/60 \times R_0$), respectively. Chosen length and velocity scales yield a Reynolds number $Re={R_0\,V_0}/{\nu}=4.2\times 10^6$ (with $\nu$ the kinematic viscosity). As reported by \cite{Susan06}, the experimental measurements were taken on a survey section located slightly below the runner outlet, where the local radius is slightly larger ($R_{\max} = 21.26$ cm). The non-dimensional radius then spans the interval $r \in [0, 1.063]$.

\begin{table}
	\scriptsize
	\centering
	\setlength{\tabcolsep}{1pt}
	\begin{tabular}{c|cccccccc|cccc}
		\% BEP &  $\Omega_0$ & $\Omega_1$ & $\Omega_2$  & $W_0$ & $W_1$ & $W_2$ & $R_1$ & $R_2$& $R_B$ & $R_C$ & $k_B$ & $k_C$ \\[2.0pt]
		
		\textbf{0.92} &~0.31765  &~-0.62888  &~2.2545  &~0.30697  &~0.01056  &~-0.31889  &~0.46643  &~0.13051  &~0.9545  &~0.1930   &~0.00432  &~0.01291 \\
		
		0.98 &~0.26675 &~-0.79994 &~3.3512 &~0.31501 & ~0.07324 &~-0.29672 &~0.36339 & ~0.09304~ & ~0.9545 & ~0.1545 & ~0.00454 & ~0.01142 \\
		
		\textbf{1.00} &~0.27113 &~-0.80310 &~3.4960 &~0.31991 & ~0.08710 &~-0.27350 &~0.37291 & ~0.08305~ & ~0.9495 & ~0.1410 & ~0.00471 & ~0.01155 \\
		
		1.03 &~0.27536  &~-0.81730  &~3.5187  &~0.32447  &~0.10618  &~-0.23545  &~0.38125  &~0.07188  &~0.0942  &~0.1260   &~0.00487   &~0.01118 \\
		
		\textbf{1.06} &~0.27419 &~-0.86579 &~3.2687 &~0.32916 & ~0.12677 &~-0.19061 &~0.37819 & ~0.06502~ & ~0.9370 & ~0.1175 & ~0.00470 & ~0.01015 \\
		
		1.11 &~0.28119  &~-0.77668  &~3.5520  &~0.31731  &~0.08308  &~-0.25254  &~0.38947  &~0.07904  &~0.9391  &~0.1176   &~0.00468   &~0.00648 \\
		
	\end{tabular}
	\caption{Flow parameters for different relative flow rates (\% BEP) with respect to the best efficiency point case. The values are the fitting parameters for (\ref{eq:vr})-(\ref{eq:vz}) and (\ref{eq:measuredk}).}
	\label{tab:baseflow_params}
\end{table}

\begin{figure}
	\includegraphics[width=\textwidth]{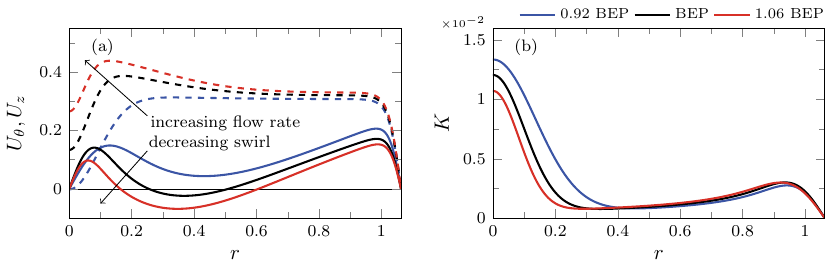}
	\caption{\label{fig:baseflow} Baseflow profiles of the three turbine operating points. (a) Azimuthal ($U_{\theta}$, solid line) and axial ($U_z$, dashed line) velocities, and (b) the corresponding turbulent kinetic energy ($K$) profiles. }
\end{figure}
  
In this study, we focus on three representative operating points: below, at, and above the BEP: $0.92$ BEP, BEP, $1.06$ BEP and further analyze the flow using linear stability analysis including a turbulent viscosity model.
Figure \ref{fig:baseflow}(a) shows the baseflow profiles of the three turbine operating points. 
To satisfy the no-slip condition at the draft tube outer wall ($r = 1.063=r_{\mathrm{max}}$), a regularization function	$H = (1 - \exp[c_{0}(r - r_{\mathrm{max}})]) / (1 + \exp[c_{0}(r - r_{\mathrm{max}})])$ is introduced, which vanishes at $r = r_{\mathrm{max}}$. The parameter $c_0$ controls the gradient of the regularization function, and a value of $c_0 = 50$ is adopted in this study. 

The three operating points differ in the intensity of the axial and azimuthal velocity components. At a flow rate below the BEP (0.92 BEP), the axial velocity profile is weakest, while the swirl velocity is stronger. As the flow rate increases to the BEP, the axial velocity increases and the swirl decreases. In the radial band $r \sim [0.25,\ 0.5]$, the swirl velocity becomes negative, indicating counter-rotation relative to the inner and outer regions. When the flow rate is increased further to 1.06 BEP, the axial velocity reaches its maximum around $r=0.1$, and the overall swirl intensity decreases, but the region of negative azimuthal velocity expands to $r \sim [0.17,\ 0.6]$.

Figure \ref{fig:baseflow}(b) shows the corresponding dimensionless turbulent kinetic energy $K$, measured and also fitted onto a continuous function defined as \citep{Susan10}
\begin{equation}
	K(r)={k_C \exp \left(-\frac{r^2}{R_C^2}\right)}+{k_B \frac{r_{\mathrm{max}}-R_B}{r_{\mathrm{max}}-r}\left[1-\exp \left(-\frac{\left(r_{\mathrm{max}}-r\right)^2}{\left(r_{\mathrm{max}}-R_B\right)^2}\right)\right]}, 
	\label{eq:measuredk}
\end{equation}
where the turbulent parameters $k_B$, $k_C$, $R_B$, and $R_C$ are provided in Table \ref{tab:baseflow_params}. The function $K$ attains its maximum in the vortex core, and reaches the highest core value at the lowest flow rate (0.92 BEP). It then decreases to a local minimum around $r \sim 0.3$, slightly increases to a local maximum near $r \sim 0.95$, and vanishes at the wall. For larger radii ($r>0.4$), all the profiles exhibit similar behaviour across the operating points. Note that the turbulent kinetic energy fitting function (\ref{eq:measuredk}) is not regularized (since it already vanishes at the wall). Only the velocity profiles are regularized.

\subsection{Governing equation and linear stability analysis (LSA)}
\label{sec:govern_eqn} 

The incompressible fluid motion is described by the velocity and pressure fields $(\mathbf{u},p)$ in cylindrical coordinates $(r, \theta, z)$ as
\begin{eqnarray}
	\frac{\partial \mathbf{u}}{\partial t}+(\mathbf{u} \cdot \nabla) \mathbf{u} & = & -\frac{1}{\rho} \nabla p + \nu \nabla^2 \mathbf{u}, \\
	\nabla \cdot \mathbf{u} & = & 0,
\end{eqnarray}
where $\rho$ is the constant density and $\nu$  the kinematic viscosity. The velocity and pressure fields are assumed to be linearly decomposed around the baseflow, described by \eqref{eq:vr}-\eqref{eq:vz}\footnote{The batchelor vortices are steady solution of the Navier-Stokes equations if the viscous diffusion of the baseflow is assumed to be exactly balanced by an additional body force, consistent with prior instability analyses of swirling jets \citep{delbende1998absolute}.}, and the infinitesimal perturbations, as
\begin{equation}
	[\mathbf{u}, p](r, \theta, z, t)  =  [\mathbf{U}_b, P](r, \theta, z) + \epsilon [\mathbf{u}', p'](r, \theta, z, t).
\end{equation}
At order $\mathcal{O}(\epsilon^1)$, the linearized Navier-Stokes equations read
\begin{eqnarray}
	\frac{\partial \mathbf{u}'}{\partial t} + \mathbf{U}_b \cdot \nabla \mathbf{u}' + \mathbf{u}' \cdot \nabla \mathbf{U}_b & = & -\frac{1}{\rho} \nabla p' + \nu \nabla^2 \mathbf{u}', \label{eq:lnse1}\\
	\nabla \cdot \mathbf{u}' & = & 0  \label{eq:lnse2}.
\end{eqnarray}

We take the view of local stability analysis, assuming that baseflow is parallel (homogeneous along $z$) and axisymmetric. The normal mode ansatz for the perturbation field reads
\begin{equation}
	\label{eq:ans}
	[\mathbf{u}', p'](r, \theta, z, t) =	[\hat{\mathbf{u}}, \rho \hat{p}](r) \mathrm{e}^{(\sigma t + \mathrm{i} m \theta + \mathrm{i} k z)} + c.c.,
\end{equation}
where $\sigma = \lambda + \mathrm{i}\omega$ is the complex frequency with the growth rate ($\lambda$) and frequency ($\omega$), $m$ and $k$ are the azimuthal and axial wavenumbers, respectively, and $c.c.$ stands for the complex conjugate.
As $\sigma\in\mathbb{C}$, $k\in\mathbb{R}$, and $m\in\mathbb{Z}$, what follows is a temporal stability analysis as opposed to spatial or spatio-temporal approaches \citep{huerre1990local}. Substituting this ansatz \eqref{eq:ans} to \eqref{eq:lnse1}--\eqref{eq:lnse2} yields 
\begin{eqnarray}
	\sigma \hat{\mathbf{u}} + \mathbf{U}_b \cdot \nabla \hat{\mathbf{u}} +\hat{\mathbf{u}} \cdot \nabla \mathbf{U}_b & = & - \nabla \hat{p} + \nu \nabla^2 \hat{\mathbf{u}}, \label{eq:lsa} \\
	\nabla \cdot \hat{\mathbf{u}} & = & 0, \label{eq:lsa2}
\end{eqnarray}
which is an eigenvalue problem for the eigenvalue $\sigma$ associated with an eigenvector $(\hat{\mathbf{u}}, \hat{p})$ where $\nabla$ replaces $\partial_\theta \to \mathrm{i}m$ and $\partial_z \to \mathrm{i}k$. Here, the pressure is normalized by the density $\rho$, and the equations are written in dimensionless form, making $\nu = 1/Re$.

Taking the turbulent interactions into account is possible by assuming that the turbulent stresses can be modelled with an eddy viscosity approach.
Following \cite{hwang2010amplification} and \cite{Viola14}, the linear stability analysis is conducted by modifying the viscous term in \eqref{eq:lsa} as  
\begin{equation}
	\nabla \cdot \left[(\nu+\nu_t(r)) \left( \nabla \hat{\mathbf{u}} + (\nabla \hat{\mathbf{u}})^T \right) \right] =
	\nu \nabla \cdot \left( \nabla \hat{\mathbf{u}} + (\nabla \hat{\mathbf{u}})^T \right)
	+ \nabla \cdot \left[\nu_t(r) \left( \nabla \hat{\mathbf{u}} + (\nabla \hat{\mathbf{u}})^T \right)\right],
	\label{eq:turbViscosity}
\end{equation}
where $\nu_t$ is the turbulent eddy viscosity and the superscript $T$ denotes transpose. The frozen eddy viscosity assumption is used, which is a simplification with respect to an approach where the turbulence model yielding $\nu_t$ would have to be linearized.

In the present study, we consider four different treatments of the eddy viscosity $\nu_{t}$ in the linear stability analysis: (a) no eddy viscosity, (b) a spatially constant eddy viscosity, (c) a mixing-length model with the radial variation of the baseflow, and (d) an eddy viscosity inferred from an experimental turbulent kinetic energy profile. These options are summarized and discussed below.

\begin{itemize}
	\setlength{\itemsep}{1ex} 
	\setlength{\labelsep}{0.5em} 
	
	\item \textbf{Baseline:} $\nu_{t}=0$.  
	This case corresponds to a purely molecular-viscosity model ($Re=4.2\times 10^6$). In the asymptotic limit $\nu \to 0$, the formulation reduces to an inviscid linear stability problem. This case provides a reference against which the effects of added turbulent diffusion can be assessed.
	
	\item \textbf{Constant eddy viscosity:} $\nu_{t}=\overline{\nu_{t}}$.\\
	A simple way to include the effect of turbulence is to replace the molecular viscosity by an effective viscosity $\nu+\nu_{t}$ with a spatially uniform turbulent contribution. We take the constant value $\overline{\nu_{t}}$ as the area-weighted average of a radially varying field,
	\begin{equation}
		\label{eq:nut_ave}
		\overline{\nu_{t}} = \frac{\int_0^{r_\mathrm{max}} \nu_{t}(r)\, r\, \mathrm{d}r}{\int_0^{r_\mathrm{max}} r\, \mathrm{d}r}.
	\end{equation}
	This choice is equivalent to performing a laminar linear stability analysis at an effective Reynolds number based on $\nu+\overline{\nu_{t}}$. The constant model provides a minimal, one-parameter correction to the baseline and is useful to test sensitivity to a uniform turbulent mixing. In the present study, the underlying field $\nu_t(r)$ is taken from the $k-\varepsilon$ model, $\nu_{t,k-\varepsilon}(r)$, defined in \eqref{eq:nu_ke}.
	
	\item \textbf{Mixing-length model:} $\nu_{t} = \nu_{t,\mathrm{ml}}(r)$.\\
	We employ a simple mixing-length model that accounts for the dominant shear contributions in the axisymmetric baseflow:
	\begin{equation}
		\label{eq:nu_ml_corrected}
		\nu_{t,\mathrm{ml}}(r) = \ell_m^2 \sqrt{\left(r\,\frac{\mathrm{d}\Omega}{\mathrm{d}r}\right)^2 + \left(\frac{\mathrm{d}U_z}{\mathrm{d}r}\right)^2},
	\end{equation}
	where $\ell_m$ is the mixing length. This formulation is based on the mean rate-of-strain tensor $S_{ij}$, as proposed by \cite{smagorinsky1963general}, such that $\nu_{t} = \ell_m^2 \sqrt{2 S_{ij} S_{ij}}$, and is commonly used in large-eddy simulations to model subgrid-scale eddies. This model is slightly more complex than the constant eddy-viscosity model, as it depends on the radial shear of the baseflow,
	resulting in a spatially varying $\nu_t$.
	For our computations, we set the non-dimensional $\ell_m = 0.025$, but we examine the sensitivity to the viscous length scale in the following sections.
	
	\item \textbf{$k\!-\!\varepsilon$ based eddy viscosity from experiment:} $\nu_{t} = \nu_{t,k\!-\!\varepsilon}(r)$.\\
	The last model is to adopt the measured turbulent kinetic energy given in \eqref{eq:measuredk}. We employ a standard $k\!-\!\varepsilon$ relation to compute the corresponding eddy viscosity $\nu_{t,k\!-\!\varepsilon}(r)$ from $\nu_t = C_\mu K^2/\varepsilon$ and $\varepsilon = C_\mu^{3/4} K^{3/2}/\ell$, which upon elimination of $\varepsilon$ gives
	\begin{equation}
		\label{eq:nu_ke}
		\nu_{t,k\!-\!\varepsilon}(r) = \ell\, C_\mu^{1/4}\, \sqrt{K(r)}.
	\end{equation}
	Here, $C_\mu = 0.09$ and the non-dimensional turbulent length scale $\ell = 0.02$ as reported by \cite{Susan10} in best agreeement with the experimental data. Compared with the previous models, this approach ties $\nu_t$ directly to measured turbulence levels and can therefore capture operating-point dependent variations in turbulent mixing.
\end{itemize}

\begin{figure}
	\centering
	\includegraphics[width=0.5\textwidth]{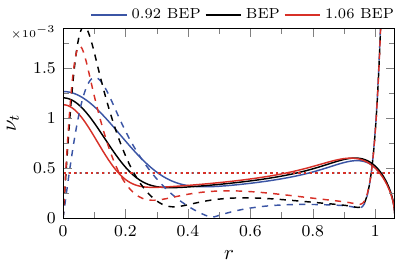}
	\caption{\label{fig:basevisc} 
			Turbulent viscosity ($\nu_t$) profiles for three turbine operating points. The profiles correspond to the constant model ($\overline{\nu_t}$, dotted lines), the mixing-length model ($\nu_{t,\mathrm{ml}}$, dashed lines), and the $k\!-\!\varepsilon$ model ($\nu_{t,k\!-\!\varepsilon}$, solid lines).}
\end{figure}

Figure~\ref{fig:basevisc} shows the turbulent viscosity profiles using the different models for the three turbine operating points. The constant eddy viscosity profiles are nearly overlapping but are not exactly equal ($\overline{\nu_{t}}_{,0.92~\mathrm{BEP}}=4.525 \times 10^{-4}$, $\overline{\nu_{t}}_{,\mathrm{BEP}}=4.509 \times 10^{-4}$, $\overline{\nu_{t}}_{,1.06~\mathrm{BEP}}=4.524 \times 10^{-4}$). The mixing-length turbulent viscosity vanishes at the axis, where the gradients of $\Omega$ and $U_z$ are zero, and at the wall, reaching values around $0.006$ due to the steep velocity gradients. This near-wall behavior is attributed to the regularization applied to enforce the no-slip condition. In contrast, the $k\!-\!\varepsilon$ turbulent viscosity attains finite values at the axis and decreases to zero at the wall. For both models, large turbulent viscosity values seem to occur in the core region, $r < 0.3$.

Eigenvalue problem \eqref{eq:lsa}--\eqref{eq:lsa2}, together with the modified diffusion term \eqref{eq:turbViscosity} accounting for the presence of $\nu_t(r)$, reads
\begin{eqnarray}
	\sigma \hat{\mathbf{u}} + \mathbf{U}_b \cdot \nabla \hat{\mathbf{u}} +\hat{\mathbf{u}} \cdot \nabla \mathbf{U}_b & = & - \nabla \hat{p} + (\nu+\nu_t) \nabla^2 \hat{\mathbf{u}} + \nabla \nu_t \left( \nabla \hat{\mathbf{u}} + (\nabla \hat{\mathbf{u}})^T \right),\\
	\nabla \cdot \hat{\mathbf{u}} & = & 0.
\end{eqnarray}
Similar equation can be obtained following the triple decomposition approach in \cite{Viola14}, as originally proposed by \cite{reynolds1972mechanics}, where the unsteady baseflow is decomposed into time-averaged baseflow, coherent fluctuation, and turbulent motion. In the present formulation, turbulence enters the eigenvalue problem only through the spatially varying eddy viscosity $\nu_t(r)$, which modifies the diffusion operator in \eqref{eq:turbViscosity}.

Expanding above equations in cylindrical coordinates, assuming $\mathbf{U}=\mathbf{U}(r)$, we obtain 
\begin{align}
	\mathcal{E}  \hat{u}_r -2\frac{U_\theta}{r} \hat{u}_\theta + \mathcal{D}_r & = -\frac{\partial \hat{p}}{\partial r} + ({\nu}+\nu_t) \left[ \mathcal{F} (\hat{u}_r) - \frac{2 \mathrm{i} m \hat{u}_\theta}{r^2} \right] + \mathcal{G}_r \frac{\partial \nu_t}{\partial r}, \label{eq:pNS1tv}\\
	\mathcal{E}  \hat{u}_\theta + \left(\frac{\partial U_\theta}{\partial r}+\frac{U_\theta}{r} \right) \hat{u}_r + \mathcal{D}_\theta &=  -\frac{\mathrm{i} m \hat{p}}{r} + ({\nu}+\nu_t) \left[ \mathcal{F} (\hat{u}_\theta) + \frac{2 \mathrm{i} m \hat{u}_r}{r^2}  \right]  + \mathcal{G}_\theta \frac{\partial \nu_t}{\partial r}, \label{eq:pNS2tv}\\
	\mathcal{E}  \hat{u}_z + \frac{\partial U_z}{\partial r} \hat{u}_r + \mathcal{D}_z &=  -\mathrm{i} k \hat{p} + ({\nu}+\nu_t) \left[ \mathcal{F} (\hat{u}_z) + \frac{\hat{u}_z}{r^2} \right] + \mathcal{G}_z \frac{\partial \nu_t}{\partial r}, \label{eq:pNS3tv}\\
	\frac{\partial \hat{u}_r}{\partial r} + \frac{\hat{u}_r}{r} +\frac{\mathrm{i} m \hat{u}_\theta}{r} +\mathrm{i} k \hat{u}_z &= 0, \label{eq:pNS4tv}
\end{align}
where
\begin{eqnarray}
	\mathcal{D}_r & = &  U_r \frac{\partial \hat{u}_r}{\partial r} + \hat{u}_r \frac{\partial U_r}{\partial r}, \quad
	\mathcal{D}_\theta  =   U_r \frac{\partial \hat{u}_\theta}{\partial r} + \frac{U_r \hat{u}_\theta}{r}, \quad
	\mathcal{D}_z  =   U_r \frac{\partial \hat{u}_z}{\partial r}, \\
	\mathcal{E} & = & \left[\sigma  + \left(\frac{U_\theta}{r} \mathrm{i} m + U_z \mathrm{i} k\right) \right], \quad
	\mathcal{F}  = \frac{\partial^2 }{\partial r^2} +\frac{1}{r} \frac{\partial}{\partial r}  -\left(\frac{m^2+1}{r^2} + k^2 \right), \label{eq:F}\\
	\mathcal{G}_r & = & 2\frac{\partial \hat{u}_r}{\partial r}, \quad
	\mathcal{G}_\theta  =  \frac{\partial \hat{u}_\theta}{\partial r} + \frac{\mathrm{i} m {\hat{u}_r} }{r}- \frac{\hat{u}_\theta}{r}, \quad
	\mathcal{G}_z  =  \frac{\partial \hat{u}_z}{\partial r} + \mathrm{i} k {\hat{u}_r} \label{eq:G}.
\end{eqnarray}

The above direct perturbation equations in cylindrical coordinates are completed with the following boundary conditions. At the wall $(r=r_{\mathrm{max}})$, a no-slip condition is imposed: $\hat{\mathbf{u}} = 0$. At the axis $(r=0)$, the boundary conditions depend on the azimuthal wavenumber $m$ and are specified following \cite{batchelor1962analysis}:
\begin{equation}
	\begin{cases} 
		\hat{u}_r = \hat{u}_\theta = 0, & m=0,\\
		\hat{u}_z = \hat{p} = 0, & {|m|}=1,\\
		\hat{u}_r = \hat{u}_\theta = \hat{u}_z = 0, & {|m|}>1.
	\end{cases}
\end{equation}

In the present local stability analysis, we restrict $k$ to be positive without the loss of generality. This is due to the following symmetry of the eigenvalue problem: if $\sigma=\lambda+\mathrm{i}\omega$ is the eigenvalue for $(m,k)$, then $\sigma^*=\lambda-\mathrm{i}\omega$ is the eigenvalue for $(-m,-k)$.

The direct problem is solved by writing it in a matrix eigenvalue problem form as
\begin{equation}
	\mathscr{L}\hat{\mathbf{q}} = \sigma \hat{\mathbf{q}},
	\label{eq:evp}
\end{equation}
where $\mathscr{L}$ is the linear operator and each of the eigenvalues $\sigma$ has a corresponding eigenvector $\hat{\mathbf{q}}=[\hat{\mathbf{u}},\hat{p}]$ and normalized so that $\left \langle \hat{\mathbf{u}}, \hat{\mathbf{u}}\right \rangle = 1$, where $\langle\cdot,\cdot\rangle$ is an inner product defined as $\left<\boldsymbol{u}_A, \boldsymbol{u}_B\right>=\int \boldsymbol{u}_A^* \boldsymbol{u}_B  ~r \mathrm{d} r$ and $^*$ denotes the complex conjugate. Each eigenvalue-eigenvector pair is called a \textit{direct mode}. If $\lambda > 0$, this mode is linearly unstable. Dynamics linearised around the baseflow will be, for a large enough time horizon, dominated by the most unstable linear mode. In this sense, this mode governs the linear stability of the system.

The adjoint perturbation equations are then obtained by computing the inner product of the direct problem with the adjoint solution $\hat{\mathbf{q}}^\dagger=[\hat{\mathbf{u}}^\dagger,\hat{p}^\dagger]$ as follows 
\begin{align}
	\mathscr{L}^{\dagger} \hat{\textbf{q}}^{\dagger} &= \sigma^* \hat{\textbf{q}}^{\dagger}. \label{eq:discrete}
\end{align}

In the discrete approach, $\mathscr{L}^{\dagger}$ is taken as the Hermitian transpose of the matrix representation of $\mathscr{L}$. 
 
Each of the adjoint eigenvalues $\sigma^* = \lambda - \mathrm{i} \omega$ has a corresponding eigenvector $\hat{\textbf{q}}^{\dagger}$ which is conveniently normalized following the biorthogonal normalization $\left \langle \hat{\mathbf{u}}^\dagger_i, \hat{\mathbf{u}}_j\right \rangle = \delta_{ij}$, where $\delta_{ij}$ is the Kronecker delta. Each adjoint eigenvalue-eigenvector pair is referred to as an \textit{adjoint mode}. Both the direct and adjoint solutions will be used in the sensitivity analysis in \S \ref{sec:sensitivity_analysis}.
	
\subsection{Wentzel–Kramers–Brillouin (WKB) analysis}
\label{sec:WKB_analysis}

The LSA introduced in the previous section computes numerically growth rates and frequencies. Under certain assumptions, the analysis can even be reduced to a purely analytical dispersion relation using WKB method. It provides a global approximation for linear differential equations in which the highest-order derivative is multiplied by a small parameter, as in the inviscid short-wavelength limit where the leading radial term takes the form \citep{Bender1978, Billant_Gallaire_2013}
\begin{equation}
		\epsilon^2 \frac{\partial^2 \psi}{\partial r^2} + \psi = 0,
\end{equation}
where $\psi$ is the flow parameter.

 In \cite{Billant_Gallaire_2013}, they reduce the Euler equation as only a function of radial velocity  $\psi=u_r \sqrt{r} / \sqrt{k^2+m^2/r^2}$, introducing a large total wavenumber limit such that $\epsilon = 1/\kappa$. The definition of the large total wavenumber $\kappa = \sqrt{k^2 + m^2}$ extended both the instability criterion of \citet{Leibovich_Stewartson_1983}, which assumes large azimuthal wavenumber, and their earlier WKB study of a purely swirling vortex at large axial wavenumber \citep{Billant_Gallaire_2005}, making the criterion applicable even for small values of $m$.
	
The governing equations are reorganized according to powers of $\kappa$, leading to an analytical dispersion relation for the complex eigenvalue $\sigma = \lambda + i \omega$, where $\lambda$ is the temporal growth rate and $\omega$ is the angular frequency. The asymptotic expression reads
\begin{align}
		{\sigma}= & -\ii \kappa \Lambda\left(r_0\right)+ \sqrt{-\Phi\left(r_0\right)} 
		- \frac{2 n+1}{2 \kappa \sqrt{2 f\left(r_0\right)}} \sqrt{\Phi^{\prime \prime}\left(r_0\right)-\frac{\Phi^{\prime}\left(r_0\right)^2}{2 \Phi\left(r_0\right)}+2 \mathrm{i} \sqrt{-\Phi\left(r_0\right)} \kappa \Lambda^{\prime \prime}\left(r_0\right)} \nonumber  \\
		& - \frac{H\left(r_0\right)}{2 \kappa f\left(r_0\right)}+O\left(\frac{\Lambda^{\prime \prime}\left(r_0\right)}{\kappa}\right) 
		- \kappa^2 {\nu}(r_0),
		\label{eq:generalomega}
\end{align}
where
\begin{align}
		\Lambda = \frac{k}{\kappa} U_z + \frac{m}{\kappa} \frac{U_\theta}{r}, \quad
		\Phi = \frac{1}{k^2 + m^2/r^2} &\left( k^2 \phi - 2 m k \frac{U_\theta U_z'}{r^2} \right), \quad \phi = 2 \frac{U_\theta}{r} \zeta, \label{eq:genom_detaildef} \\
		H = \frac{i\ \mathrm{sin} \alpha}{r^2}\left(r\zeta' - \frac{\Phi r}{U_\theta}\right) + i\ \mathrm{cos}\alpha ~r &\left(\frac{U_z'}{r}\right)', \quad
		f = \mathrm{cos} \alpha^2 + \frac{\mathrm{sin} \alpha^2}{r^2},
\end{align} 
where $k=\kappa \cos \alpha $,  $m=\kappa \sin \alpha$ with $|\alpha| \in [0,\ \pi/2]$, the axial vorticity $\zeta= (1/r) (rU_\theta)'$, the prime $'$ symbol denotes differentiation with respect to $r$, and $r_0$ is the leading-order stationary point defined by
\begin{equation}
		\left.\frac{\partial \Lambda}{\partial r}\right|_{r_0}
		=
		i\,\frac{1}{2 \kappa \sqrt{-\Phi(r_0)}}
		\left.\frac{\partial \Phi}{\partial r}\right|_{r_0}.
		\label{eq:r0def}
\end{equation}
	
The expression obtained by \citet{Billant_Gallaire_2013} is formally inviscid. The final term in (\ref{eq:generalomega}), $-\kappa^2 \nu(r_0)$, is not part of the original formulation and is added here in an \emph{ad-hoc} manner to account for dissipation, following \citet{Ok18b}, and is asymptotically valid when $\kappa^2 \gg 1$ and $\nu \ll 1$, such that $\kappa^2 \nu$ remains finite. This term may represent either molecular viscosity or turbulent eddy viscosity. Its scaling with $\kappa^2$ indicates that dissipation becomes increasingly important for short-wavelength perturbations.
	
In addition to the large total wevenumber limit, another requirement for the validity of \eqref{eq:generalomega} is that the stationary point satisfies $r_0 \gg 1/\kappa$, ensuring that the local solution around $r_0$ can be consistently matched to the inner solution near the vortex axis. When this condition is not strictly satisfied, quantitative accuracy may deteriorate. In the succeeding section, we compare the WKB predictions with the LSA results for inviscid cases to determine the parameter ranges where the WKB approximation remains accurate, and discuss how the inclusion of turbulent viscosity modifies the growth rates.
	
Furthermore, it is important to relate the WKB potential $\Phi$ to the local centrifugal instability criterion introduced by \citet{Leibovich_Stewartson_1983}. In that formulation, the generalized Rayleigh discriminant,
\begin{equation}
		\label{eq:condition}
		\tilde{\Phi}(r) = \frac{2 \Omega \Omega' r}{\Omega'^2 r^2 + U_z'^2} \left( \Omega' r \zeta + U_z'^2 \right),
\end{equation}
provides a simple, dimensionless measure of the balance between the destabilizing effect of swirl (through the angular velocity $\Omega = U_\theta/r$) and the stabilizing effect of axial shear $U_z'$. This parameter provides a sufficient condition for centrifugal instability: the flow is unstable wherever $\tilde{\Phi}(r) < 0$. 
	
In the short-wavelength limit, the WKB potential $\Phi$ (terms in \eqref{eq:genom_detaildef})
 provides a quantitative evaluation of the growth rate and frequency for given wavenumbers $k$ and $m$, while fully accounting for the radial structure of the baseflow. In contrast, $\tilde{\Phi}$ offers a simple, local measure indicating centrifugal instability. The maximum growth rate can be approximated as
\begin{equation}
		\lambda_\mathrm{max} \approx \sqrt{-\min \tilde{\Phi}(r)},
		\label{eq:max_lambda}
\end{equation}
providing a physically intuitive estimate without solving the full eigenvalue problem. This approximation corresponds to the leading-order term (2nd term) in \eqref{eq:generalomega}, as $\min \tilde{\Phi}$ occurs at the critical radius $r_0$. 
	
\subsection{Sensitivity to baseflow modifications}
\label{sec:sens_baseflow}
	
In this section, we study the structural sensitivity of the eigenvalue to the baseflow modification. Following \cite{Marquet08}, we investigate the variations $\delta \sigma$ of the complex eigenvalue $\sigma$ with respect to small-amplitude baseflow modifications $\delta \textbf{U}$. The formulation used here is identical to that of \cite{Marquet08} and is briefly outlined below. The variation $\delta \sigma$ reads 
	\begin{equation}
		\delta \sigma 
		= \langle \hat{\mathbf{u}}^\dagger, \delta \mathscr{L} \, \hat{\mathbf{u}} \rangle 
		= \Big\langle \hat{\mathbf{u}}^\dagger, \Big(\frac{\partial \mathscr{L}}{\partial \mathbf{U}} \delta \mathbf{U} \, \hat{\mathbf{u}}\Big) \Big\rangle 
		= \langle \nabla_{\mathbf{U}} \sigma, \delta \mathbf{U} \rangle,
		\label{eq:deltasigma}
	\end{equation}
which expresses the first-order change in the eigenvalue in terms of the baseflow perturbation $\delta \mathbf{U}$. Variations of the growth rate $\delta \lambda$ and frequency $\delta \omega$ are expressed as
	\begin{equation}
		\delta \lambda = 
		\left \langle \nabla_{\mathbf{U}} \lambda, \delta \mathbf{U} \right \rangle, \quad 
		\delta \omega = 
		\left \langle \nabla_{\mathbf{U}} \omega, \delta \mathbf{U} \right \rangle,
	\end{equation}
where the growth rate and frequency sensitivities can be defined as $\nabla_{\mathbf{U}} \lambda = \mathrm{Re}\{\nabla_{\mathbf{U}} \sigma\}$ and $\nabla_{\mathbf{U}} \omega = -\mathrm{Im}\{\nabla_{\mathbf{U}} \sigma\}$, respectively.
	
Expanding in cylindrical coordinates, the following sensitivities of the eigenvalue to the different baseflow velocity components are obtained:
	\begin{align}
		\nabla_{U_r} \sigma &= \underbrace{-\left( \frac{\partial \hat{u}_r^*}{\partial r} \hat{u}_r^\dagger + \frac{\partial \hat{u}_\theta^*}{\partial r} \hat{u}_\theta^\dagger + \frac{\partial \hat{u}_z^*}{\partial r} \hat{u}_z^\dagger \right)}_{\nabla_{U_r, \mathrm{trans}} \sigma}
		+ \underbrace{ \frac{1}{r} \left[ \frac{\partial}{\partial r}\!\left( r \hat{u}_r^* \hat{u}_r^\dagger \right) - \hat{u}_\theta^* \hat{u}_\theta^\dagger \right]}_{\nabla_{U_r, \mathrm{prod}} \sigma}, \label{eq:sens_base1}\\[6pt]
		\nabla_{U_\theta} \sigma &= 
		\underbrace{\frac{\mathrm{i} m}{r} \left( 
			\hat{u}_r^* \hat{u}_r^\dagger 
			+ \hat{u}_\theta^* \hat{u}_\theta^\dagger 
			+ \hat{u}_z^* \hat{u}_z^\dagger \right)}_{\nabla_{U_\theta, \mathrm{trans}} \sigma}
		+ \underbrace{\frac{1}{r} \left[ 
			\frac{\partial}{\partial r}\!\left( r \hat{u}_r^* \hat{u}_\theta^\dagger \right) 
			- \hat{u}_r^* \hat{u}_\theta^\dagger + 2 \hat{u}_\theta^* \hat{u}_r^\dagger \right]}_{\nabla_{U_\theta, \mathrm{prod}} \sigma}, \label{eq:sens_base2}\\[6pt]
		\nabla_{U_z} \sigma &= 
		\underbrace{\mathrm{i} k \left( 
			\hat{u}_r^* \hat{u}_r^\dagger
			+ \hat{u}_\theta^* \hat{u}_\theta^\dagger 
			+ \hat{u}_z^* \hat{u}_z^\dagger \right)}_{\nabla_{U_z, \mathrm{trans}} \sigma}
		+ \underbrace{\frac{1}{r} \frac{\partial}{\partial r}\!\left( r \hat{u}_r^* \hat{u}_z^\dagger \right)}_{\nabla_{U_z, \mathrm{prod}} \sigma} \label{eq:sens_base3}.
	\end{align}
Here, the sensitivity to baseflow modifications is decomposed into contributions from transport ($\nabla_{U, \mathrm{trans}} \sigma$) and production ($\nabla_{U, \mathrm{prod}} \sigma$). These originate from the transport of perturbations by the baseflow ($\nabla \hat{\mathbf{u}} \cdot \mathbf{U}_b$) and from the production of perturbations by the baseflow ($\nabla \mathbf{U}_b \cdot \hat{\mathbf{u}}$), respectively. Similar interpretation is given by \cite{Marquet08}. Note that, even though $U_r = 0$ in the baseflow, the sensitivity $\nabla_{U_r} \sigma$ remains non-zero due to the possibility of introducing perturbations in the radial component of the baseflow velocity. In a more general setting where $\mathbf{U} = \mathbf{U}(r,\theta,z)$, additional terms involving $\partial/\partial \theta$ and $\partial/\partial z$ would explicitly appear in the linearized operator, reflecting the influence of azimuthal and axial variations in the baseflow structure.
	
\subsection{Sensitivity to turbulent viscosity modifications}
\label{sec:sens_turb_viscosity}
In this section, we derive the sensitivity of the eigenvalues to turbulent viscosity modifications. This is motivated by the work of \cite{Viola14} who reported that including turbulent viscosity (even as a form of constant eddy viscosity) can dramatically alter the eigenvalue spectrum and is able to isolate the experimentally observed dominant mode in a wind-turbine wake. This demonstrates the importance of including the effects of turbulent viscosity. Although straightforward, the sensitivity of the eigenvalues to turbulent viscosity variations has not yet been explored. Here, we derive the eigenvalue sensitivity to turbulent viscosity modifications and quantify their impact on stability.
	
As seen in (\ref{eq:turbViscosity}), the only term in $\mathscr{L}\hat{\mathbf{u}}$ that depends on the turbulent viscosity $\nu_t$ is $\nabla \cdot \left[ \nu_t(r) \left( \nabla + \nabla^T \right) \hat{\mathbf{u}} \right]$. Accordingly, similar to \eqref{eq:deltasigma}, the sensitivity functional can be expressed as
\begin{equation}
		\delta \sigma 
		= \big\langle \hat{\mathbf{u}}^\dagger, \delta \mathscr{L}\,\hat{\mathbf{u}} \big\rangle
		= \Big\langle \hat{\mathbf{u}}^\dagger, 
		\Big(\frac{\partial \mathscr{L}}{\partial \nu_t}\,\delta\nu_t\Big)\hat{\mathbf{u}} \Big\rangle
		= \big\langle \nabla_{\nu_t} \sigma, \delta \nu_t \big\rangle,
		\label{eq:sensitivityNu}
\end{equation}
where
\begin{equation}
		\delta \mathscr{L} \hat{\mathbf{u}} = \nabla \cdot \left[ \delta \nu_t(r) \left( \nabla \hat{\mathbf{u}} + (\nabla \hat{\mathbf{u}})^T \right) \right].
\end{equation}
	
From the definition and integration by parts, we have
\begin{align}
		\left\langle \hat{\mathbf{u}}^\dagger, \delta \mathscr{L} \hat{\mathbf{u}} \right\rangle 
		&= \int_\Omega \hat{\mathbf{u}}^{*\dagger} \cdot \nabla \cdot \Big[ \delta \nu_t(r) \big( \nabla \hat{\mathbf{u}} + (\nabla \hat{\mathbf{u}})^T \big) \Big] \, \mathrm{d}\Omega \nonumber \\
		&= - \int_\Omega \left( \nabla \hat{\mathbf{u}}^{*\dagger} \right) : \Big[ \delta \nu_t(r) \big( \nabla \hat{\mathbf{u}} + (\nabla \hat{\mathbf{u}})^T \big) \Big] \, \mathrm{d}\Omega \nonumber \\
		&+ \oint_{\partial \Omega} \hat{\mathbf{u}}^{*\dagger} \cdot \Big[ \delta \nu_t \big( \nabla \hat{\mathbf{u}} + (\nabla \hat{\mathbf{u}})^T \big) \mathbf{n} \Big] \, \mathrm{d}\Gamma,
\end{align}
where $:$ denotes the Frobenius inner product. The boundary term vanishes since $\hat{\mathbf{u}}^\dagger = 0$ and $\delta \nu_t = 0$ on $\Gamma$.
	
Defining $\mathbf{S} \equiv \nabla \hat{\mathbf{u}} + (\nabla \hat{\mathbf{u}})^T$, the sensitivity kernel with respect to the turbulent viscosity becomes
\begin{equation}
		\nabla_{\nu_t} \sigma = - \left( \nabla \hat{\mathbf{u}}^\dagger \right) : \mathbf{S}^* 
		= -\frac{1}{2} \left( \mathbf{S}^\dagger : \mathbf{S} \right)^*.
\end{equation}
Note that this expression is general, and is valid in any spatial dimension and coordinate system.

The expanded form of the turbulent viscosity sensitivity in cylindrical coordinates can be expressed as
\begin{align}
		\nabla_{\nu_t} \sigma 
		= ~ &C_r^* \hat{u}_r^{\dagger} - \frac{2}{r} \frac{\partial}{\partial r} \Big( r \hat{u}_r^* \hat{u}_r^{\dagger} \Big) \nonumber \\
		+ ~&C_\theta^* \hat{u}_\theta^{\dagger} - \frac{1}{r} \Bigg[ \frac{\partial}{\partial r} \Big( r \frac{\partial \hat{u}_\theta^*}{\partial r} \hat{u}_\theta^{\dagger} \Big) 
		- \mathrm{i} m \frac{\partial}{\partial r} \big( \hat{u}_r^* \hat{u}_\theta^{\dagger} \big) 
		- \frac{\partial}{\partial r} \big( \hat{u}_\theta^* \hat{u}_\theta^{\dagger} \big) \Bigg] \nonumber \\
		+ ~&C_z^* \hat{u}_z^{\dagger} - \frac{1}{r} \Bigg[ \frac{\partial}{\partial r} \Big( r \frac{\partial \hat{u}_z^*}{\partial r} \hat{u}_z^{\dagger} \Big) 
		- \mathrm{i} k \frac{\partial}{\partial r} \big( r \hat{u}_r^* \hat{u}_z^{\dagger} \big) \Bigg],
\end{align}
where
\begin{equation}
		C_r = \mathcal{F} (\hat{u}_r) - \frac{2 \mathrm{i} m \hat{u}_\theta}{r^2},\quad
		C_\theta = \mathcal{F} (\hat{u}_\theta) + \frac{2 \mathrm{i} m \hat{u}_r}{r^2},\quad
		C_z = \mathcal{F} (\hat{u}_z) + \frac{\hat{u}_z}{r^2}.
\end{equation}
	
This sensitivity of the eigenvalue to turbulent viscosity can be further decomposed into the sensitivity to modifications of the turbulent diffusion $\nabla_{\nu_t, \mathrm{diff}} \sigma$ and sensitivity to modifications of the gradient of the turbulent viscosity $\nabla_{\nu_t, \mathrm{grad}} \sigma$, which can be expressed as
\begin{align}
		\nabla_{\nu_t, \mathrm{diff}} \sigma &= C_r^* \hat{u}_r^{\dagger} + C_\theta^* \hat{u}_\theta^{\dagger} + C_z^* \hat{u}_z^{\dagger}, \\
		\nabla_{\nu_t, \mathrm{grad}} \sigma &= -\frac{2}{r} \frac{\partial}{\partial r} \Big( r \hat{u}_r^* \hat{u}_r^{\dagger} \Big) 
		- \frac{1}{r} \Bigg[ \frac{\partial}{\partial r} \Big( r \frac{\partial \hat{u}_\theta^*}{\partial r} \hat{u}_\theta^{\dagger} \Big) 
		- \mathrm{i} m \frac{\partial}{\partial r} \big( \hat{u}_r^* \hat{u}_\theta^{\dagger} \big) 
		- \frac{\partial}{\partial r} \big( \hat{u}_\theta^* \hat{u}_\theta^{\dagger} \big) \Bigg] \nonumber \\
		&\quad - \frac{1}{r} \Bigg[ \frac{\partial}{\partial r} \Big( r \frac{\partial \hat{u}_z^*}{\partial r} \hat{u}_z^{\dagger} \Big) 
		- \mathrm{i} k \frac{\partial}{\partial r} \big( r \hat{u}_r^* \hat{u}_z^{\dagger} \big) \Bigg].
\end{align}
	
Here, $\nabla_{\nu_t, \mathrm{diff}} \sigma$ originates from the turbulent viscous diffusion term which is directly proportional to the change $\delta \nu_t$, i.e., $\delta\nu_t(r) \nabla \cdot \left( \nabla + \nabla^T \right) \hat{\mathbf{u}}$, while $\nabla_{\nu_t, \mathrm{grad}} \sigma$ arises from the interaction of the gradient of turbulent viscosity with the velocity gradients, $\nabla \delta\nu_t(r) \cdot \left( \nabla + \nabla^T \right) \hat{\mathbf{u}}$.
	
\section{Numerical methods}
\label{sec:numerical}
The one-dimensional eigenvalue problems \eqref{eq:pNS1tv}-\eqref{eq:pNS4tv} are solved with finite element method using FreeFEM++ software \citep{hecht2012new}
 and the space is discretised by using \textit{segment} function. The mesh, consisting of 2000 elements, is non-uniform with the element centers distributed following $r = x^{1.2}$, where $x$ is a uniform subdivision of the unit interval with the same number of elements. Clustering the element near the axis ensure accurate resolution of steep gradients in this region. The velocity and pressure fields are discretized on the one-dimensional mesh using Lagrange finite elements of order $P_2$ and $P_1$, respectively. This ensures a consistent 1D finite-element formulation while retaining the appropriate polynomial orders for velocity and pressure. At $r=0$, \eqref{eq:vt} would pose numerical problems due to division by $r$. To avoid this difficulty, the azimuthal velocity profile given in \eqref{eq:vt} is not evaluated pointwise on the mesh but rather, an auxiliary variational problem $\int U_{\theta}\, v\,r\,dr=\int \widetilde{U_{\theta}}\, v\,r\,dr$ is solved for the discretized field $\widetilde{U_{\theta}}$ with a test function $v$. As the quadrature points used in evaluating these integrals are never on the domain boundary, the terms $1/r$ do not pose a difficulty. The same strategy is adopted later in the computation of the turbulent kinetic energy and sensitivity kernels, where factors of $1/(r_{\mathrm{max}}-r)$ and $1/r$ appear and must be handled.
	
Spatially discretized eigenvalue problems of the form \eqref{eq:evp} are solved using Implicitly Restarted Arnoldi Method, as implemented in ARPACK library \citep{arpack} in FreeFEM++. A subset of eigenvalues of sparse generalized eigenvalue problem in the vicinity of a complex shift is identified using the shift-invert method. Adjoint eigenvectors are computed following the discrete approach \eqref{eq:discrete} by solving the eigenvalue problem using the Hermitian transpose of the matrix representation of the linear operator. This approach guarantees biorthogonality of direct and adjoint vectors and the same set of eigenvalues for both direct and adjoint problems up to machine precision.

Linear stability results obtained with FreeFEM++ are validated against the Chebyshev pseudo-spectral collocation method \citep{antkowiak2004transient, Ok15a} (Appendix \ref{sec:appen_freefem_validation}) and the mesh convergence study is detailed in Appendix \ref{sec:appen_grid_study}.

\section{Results and discussion}
\label{sec:results}
We present the results of the linear stability analysis (\S \ref{sec:stability_analysis}) and the WKB stability analysis (\S \ref{sec:WKB_analysis_results}). 
The sensitivity of growth rates and frequencies to modifications in baseflow and turbulent viscosity is discussed in \S \ref{sec:sensitivity_analysis}, with the corresponding sensitivity over the axial wavenumber spectrum presented in \S \ref{sec:sensitivity_analysis_extended}. We also explore the effect of a reduced turbulent length scale on flow stability (\S \ref{sec:lower_turb_length}) and consider the stability of the swirling baseflow across the full range of baseflow conditions in \S \ref{sec:baseflow_discharge_coeff}.
\subsection{Linear stability analysis}
\label{sec:stability_analysis}

\begin{figure}
	\centering
	\begin{subfigure}{\textwidth}
		\includegraphics[width=\textwidth]{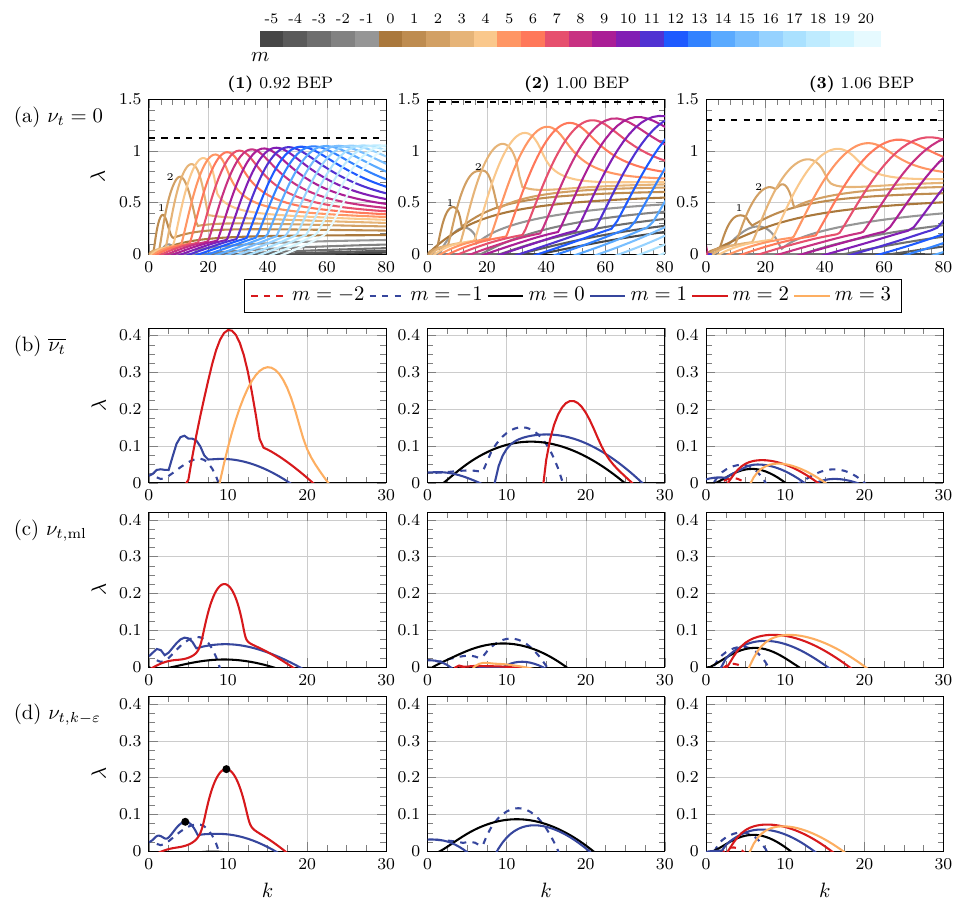}
	\end{subfigure}
	\caption{\label{fig:specturm} Growth rates $\lambda$ as functions of the axial wavenumber $k$, with each branch corresponding to a different azimuthal wavenumber $m$. The figure is organized in a $4 \times 3$ grid: rows represent different turbulence models: (a) $\nu_t = 0$, (b) $\overline{\nu_{t}}$, (c) $\nu_{t, \mathrm{ml}}$, and (d) $\nu_{t, k-\varepsilon}$, while the columns correspond to operating conditions: (1) $0.92$ BEP, (2) BEP, and (3) $1.06$ BEP. The dashed lines correspond to the predicted maximum growth rate defined in \eqref{eq:max_lambda}.}
\end{figure}

Figure~\ref{fig:specturm} presents the results of the linear stability analysis performed for four turbulence models: (a) null eddy viscosity, $\nu_t = 0$, (b) a spatially averaged turbulent viscosity, $\overline{\nu_{t}}$, defined in \eqref{eq:nut_ave}, (c) the mixing-length formulation, $\nu_{t,\mathrm{ml}}$, given in~\eqref{eq:nu_ml_corrected}, and (d) the $k-\varepsilon$ based turbulent viscosity, $\nu_{t,k-\varepsilon}$, defined in~\eqref{eq:nu_ke}.
Each column corresponds to a different operating condition: (1) $0.92~\mathrm{BEP}$, (2) BEP, and (3) $1.06~\mathrm{BEP}$. For each configuration, the growth rate is plotted as a function of the axial wavenumber $k$ for multiple azimuthal wavenumbers $m$ and the data were generated with an increment of $\Delta k = 0.1$. In figure \ref{fig:specturm}a, we computed modes only up to $m=20$. The instability continue existing for larger $m$ with similar pattern as in smaller $m$ but shifted to larger $k$. When turbulence effects are neglected as shown in figure \ref{fig:specturm}({1}a-{3a}) (i.e. $\nu_t = 0$), the growth rates reaches to a local maximum with increasing $k$ from zero then decreases to a plateau. It does not yield a clearly dominant mode which is typical of centrifugal instability in the inviscid limit, where the maximum growth rate and the most amplified wavenumber occur as $k \to \infty$ \citep{Billant_Gallaire_2005,Ok16a}. However, as illustrated in rows (b-d) of the same figure, the inclusion of turbulent diffusion through $\nu_t$ leads to a pronounced collapse of the growth rate curves, by roughly an order of magnitude, enabling the identification of dominant unstable modes. Moreover, only small wavenumber components remain unstable $|m|=0,1,2,3$, and higher-wavenumber modes are damped more strongly probably due to the diffusion term, which scales with $k^2$ and $m^2$ (see $\mathcal{F}$ in \eqref{eq:F}).
 This agrees with experimental evidence where modes $|m|=0,1,2$ are observed in the turbulent regime. These results show how turbulent diffusion strongly  damps our system, establishing its importance in our stability analysis.

Figures~\ref{fig:specturm}(b) show the growth rate spectra obtained when a spatially averaged, constant turbulent viscosity, $\overline{\nu}_{t}$, is used. This case is equivalent to performing a linear stability analysis at a reduced effective Reynolds number. Even this simple treatment introduces noticeable damping of the higher azimuthal modes. However, at $0.92\,\mathrm{BEP}$, very large growth rates are  observed for the $m=2$ and $m=3$ modes, indicating that a uniform eddy viscosity is less effective in stabilizing the most amplified structures in this operating regime.

Figures~\ref{fig:specturm}(c) shows the mixing-length model, $\nu_{t,\mathrm{ml}}$, given in~\eqref{eq:nu_ml_corrected}. Although this model does not rely on turbulent kinetic energy measurements, it does depend on the prescribed mixing length, $\ell_m$, which acts as its primary calibration parameter. As shown in figure~\ref{fig:basevisc}, the resulting $\nu_t(r)$ profiles differ from those obtained using the $k-\varepsilon$ formula, making it particularly interesting to assess their impact on the stability characteristics. The corresponding spectra, exhibit significantly stronger damping than the spatially averaged case. At 0.92 BEP, the spectrum is clearly dominated by $m=2$ mode. At BEP, the dominant modes shift to $m=0$ and $m=-1$, while at 1.06 BEP, the largest growth rates occur for $m=2$ and $m=3$, indicating a systematic redistribution of modal preference with operating condition.

Finally, figures~\ref{fig:specturm}(d) present the growth rate spectra computed using the turbulent viscosity $\nu_{t,k-\varepsilon}$, \eqref{eq:nu_ke}. In contrast to the mixing-length model, this formulation is more realistic because it incorporates experimentally measured turbulent diffusion into the linear stability analysis. The dominant trends, however, are qualitatively similar to those obtained using the mixing-length viscosity, only subtle differences are observed. In particular, at $0.92\,\mathrm{BEP}$, the $m=0$ mode is fully stabilized, while at BEP the higher-$m$ modes, specifically $m=2$ and $m=3$, are also strongly damped, resulting in an overall more strongly suppressed spectrum. This indicates that the BEP operating condition more effectively suppresses the growth of higher azimuthal modes compared with part-load and overload conditions.
\begin{figure}
	\centering
	\includegraphics[width=0.5\textwidth]{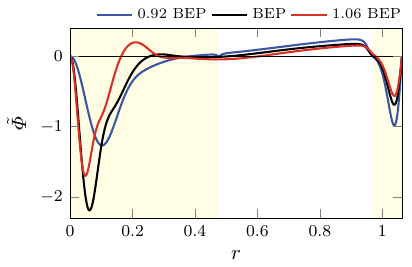}%
	\caption{\label{fig:Generalized_Rayleigh_discriminant} Leibovich \& Stewartson instability criteria  defined in \eqref{eq:condition} for the different baseflow conditions. The region in which $\tilde{\Phi} < 0$ at 0.92 BEP is highlighted in yellow.}
\end{figure}
\begin{figure}
	\includegraphics[width=\textwidth]{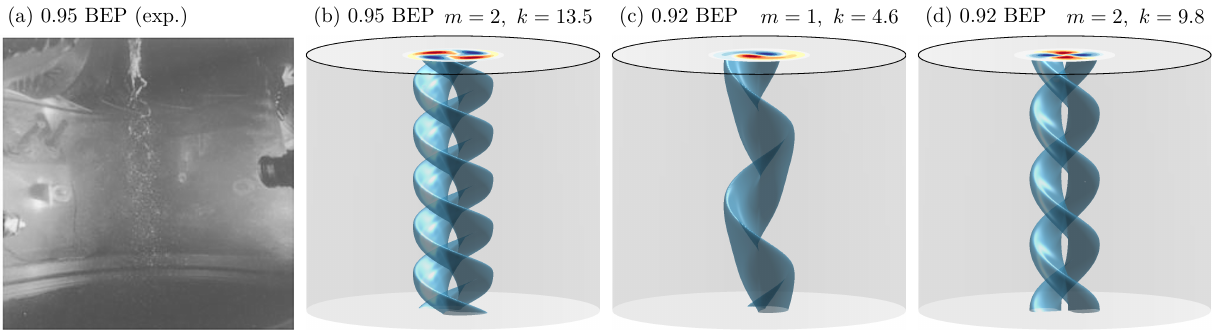}%
	\caption{\label{fig:discussion} (a) Experimental observation of air bubble visualized helical vortices at the downstream of Francis turbine at 0.95 BEP \citep{Jacob93,Muri02}, and numerical three-dimensional reconstruction (for a fixed axial length) of axial vorticities for (b) 0.95 BEP (matched with exp.): $m=2, k=13.5$, and for 0.92 BEP: (c) $m=1, k=4.6$ and (d) $m=2, k=9.8$ corresponding to the most unstable modes marked with $\bullet$ in figure \ref{fig:specturm}(1d).}
\end{figure}

It is interesting that the simple mixing-length model yields stability characteristics comparable to those obtained using the experimentally measured $k-\varepsilon$ turbulent viscosity, despite not relying on measured turbulence quantities, only through the mean flow information. This suggests that the stability behavior can be accurately predicted provided that the radial distribution of $\nu_t(r)$ is included. In contrast, the uniform viscosity $\bar{\nu}_{t}$ does not include this spatial structure, allowing higher azimuthal modes persist (e.g. $m=3$ at $0.92\,\mathrm{BEP}$), which is not observed when spatially varying turbulence models are employed. In this sense, even without measured turbulence quantities, using only the measured mean flow to construct a representative $\nu_t(r)$ (with a good approximation of $l_m$) is sufficient to recover the key effect of turbulent diffusion and obtain relevant LSA predictions for centrifugal instability.

In figures~\ref{fig:specturm}(1a-3a), the dashed lines indicate the maximum growth rates predicted using the simple local criterion $\tilde{\Phi}$, as given in \eqref{eq:max_lambda}. As observed, this inviscid criterion captures the maximum growth rates reasonably well for all operating conditions in the $\nu_t=0$ case. The Leibovich \& Stewartson instability criteria \eqref{eq:condition} for the different baseflow profiles are then shown in figure \ref{fig:Generalized_Rayleigh_discriminant}. In the inviscid limit, a sufficient condition for instability is that $\tilde{\Phi}<0$ for some radius $r$. The region satisfying this condition at 0.92 BEP is highlighted in yellow and the unstable modes are expected to be localized within this region. This concentration of the negative $\tilde{\Phi}$ region near the axis indicates a centrifugally unstable vortex. A possibly unstable region is also identified near the wall due to a strong gradient of the baseflow velocity imposed by the regularisation function.

\begin{figure}
	\includegraphics[width=\textwidth]{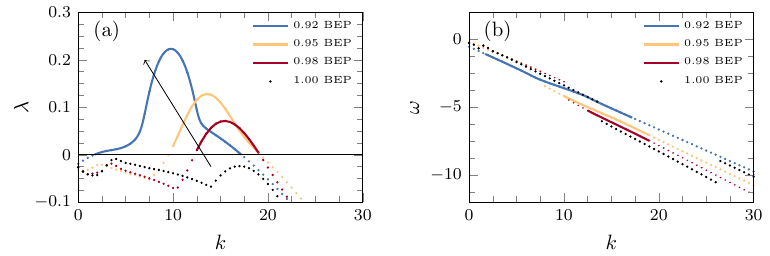}%
	\caption{\label{fig:m2_evolution} (a) Growth rate and (b) frequency plots of the $m=2$ mode under different part-load conditions and at the BEP. Solid lines indicate positive growth rates (and their frequencies), while dotted markers indicate negative growth rates (and their frequencies).}
\end{figure}

Figure~\ref{fig:discussion}(a) shows an experimental snapshot of the vortex rope at the $0.95\,\mathrm{BEP}$ operating condition \citep{Jacob93,Muri02}. A center concentrated multi-helical flow structure is observed  by the injection of air.  
 The three-dimensional reconstructions of axial vorticity of the most unstable modes are presented for a fixed axial length in figures~\ref{fig:discussion}(b-d). Specifically, figure~\ref{fig:discussion}(b) shows the reconstruction at $0.95\,\mathrm{BEP}$ for the most unstable mode $m=2$, $k=13.5$, that matches qualitatively with the experimental observation. Figures~\ref{fig:discussion}(c) and (d) correspond to $0.92\,\mathrm{BEP}$ for the most unstable modes $m=1$, $k=4.6$ and $m=2$, $k=9.8$, respectively, as identified in figure~\ref{fig:specturm}(1d) by an additional circular marker ($\bullet$). The reconstructed helical modes capture the qualitative features observed experimentally, including the azimuthal modulation and core structure associated with the dominant unstable modes.

Figure \ref{fig:m2_evolution} provides a detailed view of the growth rate and frequency evolution of the $m = 2$ mode across different part-load operating conditions. The frequency and growth rate curves versus $k$ are sometimes not smooth because only the most unstable mode is collected regardless of the instability branches (detailed in \S \ref{sec:lower_turb_length}). At the BEP, $m=2$ mode is fully stable. As the flow rate is reduced, it progressively becomes unstable, with a maximum growth rate attained at the point farthest from BEP. On the other hand, the frequency remains similar over the flow rate exhibiting a linear dependency of $k$. 

\begin{figure}
	\includegraphics[width=\textwidth]{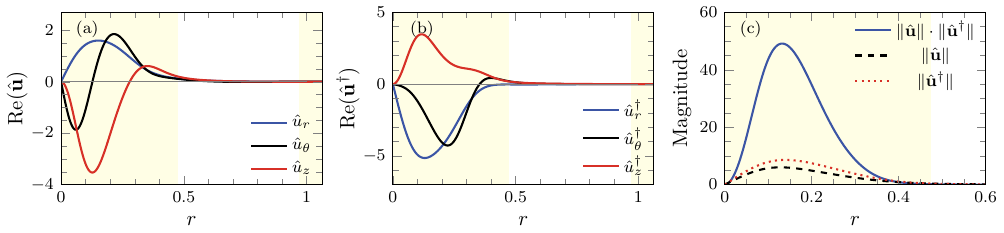}%
	\caption{\label{fig:direct_adjoint_mode} Real part of the leading (a) direct and (b) adjoint eigenvectors  for $m = 2$ and $k = 9.8$ at 0.92 BEP using $\nu_{t,k-\varepsilon}$ and (c) the wavemaker region, defined as the product of the magnitudes of the direct and adjoint modes, $\|\hat{\textbf{u}}\| \cdot \|\hat{\textbf{u}}^\dagger\|$.
	The shaded region indicates where $\tilde{\Phi}<0$.}
\end{figure}

 Figure \ref{fig:direct_adjoint_mode}(a) shows the leading real part of the direct eigenmode at $m=2$ and $k=9.8$ for 0.92 BEP, using the $k-\varepsilon$ turbulent viscosity profile. The mode is localized near the axis and within the instability region predicted by $\tilde{\Phi}<0$, highlighted by a yellow shaded region. The corresponding real adjoint mode is shown in figure \ref{fig:direct_adjoint_mode}(b). Like the direct mode, it is concentrated near the axis and exhibits similar spatial oscillations. Both the direct and adjoint modes are localized in the same region. This is in contrast to what is observed in open shear flows, where spatial support of the direct and adjoint eigenmodes is generally different (e.g. the classical flow over a cylinder, where the direct and adjoint modes are typically localized downstream and upstream respectively \citep{Marquet08}) which leads to strong non-normality.
\cite{giannetti2007structural} introduced the concept of a \textit{wavemaker} region, and \cite{Marquet08} proposed a simplified formulation using the product of the magnitudes of the direct and adjoint velocities, $\|\hat{\textbf{u}}\| \cdot \|\hat{{\bm u}}^\dagger\|$, where $\|{\textbf{u}}\|=\sqrt{u_r^*\,u_r+u_{\theta}^*\,u_{\theta}+u_z^*\,u_z} $. This quantity is shown in figure \ref{fig:direct_adjoint_mode}(c), and its distribution exhibits a bell-shaped profile peaking at $r \sim 0.15$, identifying the region where the instability is intrinsically generated and sustained. This can also be interpreted as the region of highest sensitivity to local feedback of the leading mode. For comparison, the norms of the direct $\|\hat{\textbf{u}}\|$ and adjoint $\|\hat{\textbf{u}}^\dagger\|$ eigenmodes are also plotted, showing that both modes are concentrated in the same region. The result is reminiscent, in a one-dimensional sense, of the \textit{wavemaker} region identified for the first instability of the cylinder wake (see \cite{Marquet08}). 
In \S \ref{sec:sensitivity_analysis}, we quantify this sensitivity specifically by evaluating how localized modifications of the baseflow in this region affect the eigenvalue of the leading mode.

\subsection{WKB analysis}
\label{sec:WKB_analysis_results}

\begin{figure}
	\centering
	\includegraphics[width=\textwidth]{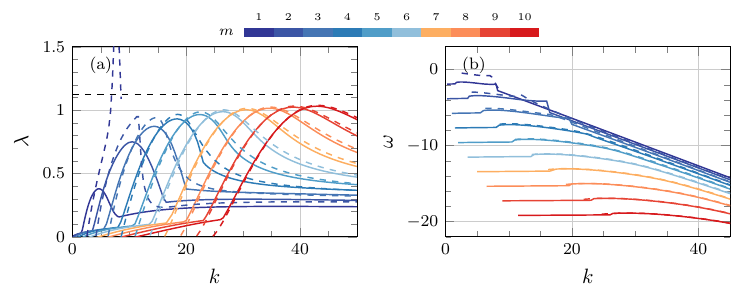}
	\caption{\label{fig:WKB_m2_compare} Comparisons between WKB (\ref{eq:generalomega}) (dashed lines) and LSA (solid lines) results: (a) growth rate and (b) frequency for the inviscid case ($\nu_{t}=0$) at 0.92 BEP. The black dashed line in (a) corresponds to the maximum growth rate predicted computed from \eqref{eq:max_lambda}.}
\end{figure}

\begin{figure}
	\centering
	\includegraphics[width=\textwidth]{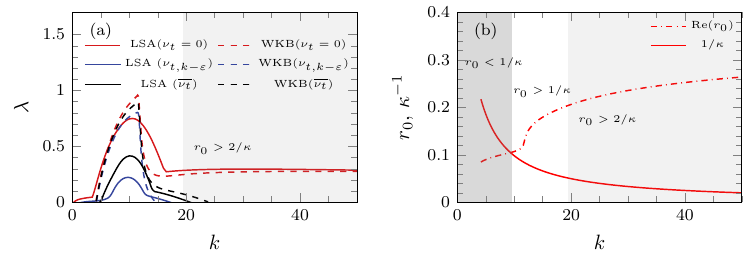}
	\caption{\label{fig:WKB_m2_comparem2} Growth rate comparisons between WKB and LSA for $m=2$ (a) for inviscid limit and viscous cases, and (b) the stationary radii and the order $1/\kappa$. }
\end{figure}

In this section, we discuss the results obtained from the WKB analysis introduced in \S\ref{sec:WKB_analysis}. Figure~\ref{fig:WKB_m2_compare} compares the spectra computed from the WKB formulation \eqref{eq:generalomega} with those obtained from the linear stability analysis, for the case $\nu_t = 0$ at 0.92 BEP. The WKB approach provides good predictions of both the growth rate and the frequency, particularly for higher azimuthal wavenumbers. For small wavenumbers $m \leq 3$, the WKB analysis begins to considerably overpredict the maximum growth rate, and the agreement deteriorates further for $m = 1$, where the prediction becomes unreliable. This behavior may due to the breakdown of the WKB validity condition, $\kappa \gg 1$. In contrast, the predicted frequencies remain in reasonable agreement with the LSA results even for $m = 1$. This can be explained by the leading-order term in \eqref{eq:generalomega}, $\omega \sim k\, U_z(r_0) + m\, U_{\theta}(r_0)$, which, for a fixed azimuthal wavenumber $m$, is essentially proportional to $k$. This explains the nearly linear dependence of the frequency on $k$ observed in figure~\ref{fig:WKB_m2_compare}(b). To further investigate the discrepancies near the maximum growth rate, and to assess the behavior of the WKB prediction in the presence of turbulent viscosity, figure~\ref{fig:WKB_m2_comparem2} focuses on the $m = 2$ case.


In figure~\ref{fig:WKB_m2_comparem2}(a), the WKB prediction is compared with the inviscid limit of \eqref{eq:generalomega} (without the viscous correction) and with the full linear stability analysis in the almost inviscid limit ($Re = 10^{6}$). Even in the inviscid case, the maximum growth rate predicted by WKB ($\lambda_{\text{WKB},\max} = 0.949$) already exceeds the value obtained from LSA ($\lambda_{\text{LSA},\max} = 0.746$). As previously discussed, this discrepancy stems from the assumptions underlying the generalized dispersion relation, namely $\kappa \gg 1$ and $r_0 \gg 1/\kappa$ (see \S\ref{sec:WKB_analysis}).

Figure~\ref{fig:WKB_m2_comparem2}(b) further illustrates this point by displaying the location of the stationary radius $r_0$ relative to $1/\kappa$. For $m = 2$ and $k < 9.6$, one finds $r_0 < 1/\kappa$, which violates the asymptotic assumption. The region where $r_0$ becomes large enough than $1/\kappa$ is also indicated (we chose the limit as $r_0 > 2/\kappa$) where the agreement between the WKB and LSA predictions in the inviscid limit improves significantly. Therefore, the overestimation of the growth rate at small $k$ can be directly attributed to the breakdown of the WKB assumptions in this regime. In such cases, one could expand the equation of motion using the multiscale matching approaches proposed by \cite{LE_DIZES_FABRE_2007,Fabre_LeDizes_2008}, however, this is beyond the scope of the present study.

When turbulent viscosity $\nu_t$ is included, a substantial reduction of the maximum growth rate is observed in the LSA results, yielding $\lambda_{\text{LSA},\nu_t,\max} = 0.223$. In contrast, the WKB prediction evaluated with $\nu_{t, k-\varepsilon}(r_0)$ shows only a minor change compared to the inviscid case, again showing its tendency to overestimate the peak growth rate. It should also be noted that the \textit{ad hoc} term in \eqref{eq:generalomega}, $-\kappa^2 \nu_T(r_0)$, introduces the effect of turbulent viscosity only at the single location $r_0$, and thus does not account for its full spatial distribution. Nevertheless, the WKB analysis correctly captures the viscous cut-off (beyond which the flow becomes stable), the most amplified wavenumber $k_{\mathrm{max}}$ for which $\lambda(k_{\mathrm{max}}) = \lambda_{\mathrm{max}}$, and the overall decay of the growth rate at large $k$. These observations further motivate a detailed investigation of the sensitivity of the eigenvalue to variations in $\nu_t$, which is the focus of the following section.

\subsection{Sensitivity analysis}
\label{sec:sensitivity_analysis}

\subsubsection{Sensitivity to baseflow modifications}
\label{sec:sensitivity_analysis_baseflow}

\begin{figure}
	\includegraphics[width=\textwidth]{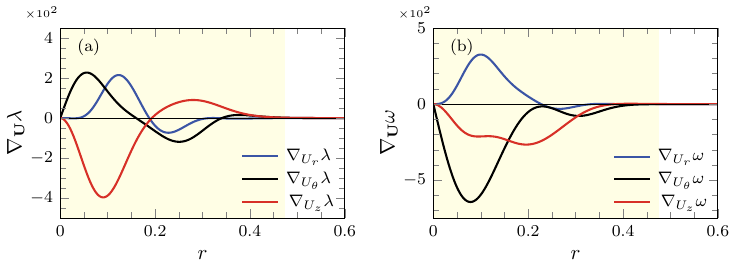}%
	\caption{\label{fig:sens_baseflow} Sensitivity to baseflow modifications of the leading eigenvalue ($\nabla_{\tiny\textbf{U}} \sigma = (\nabla_{U_r} \sigma, \nabla_{U_\theta} \sigma, \nabla_{U_z} \sigma)$) at $m=2$ and $k=9.8$ for 0.92 BEP using $\nu_{t,k-\varepsilon}$. (a) Growth rate $\nabla_{\tiny\textbf{U}} \lambda$ and (b) frequency $\nabla_{\tiny\textbf{U}} \omega$ sensitivity to baseflow modifications as a function of $r$.}
\end{figure}

We now consider the sensitivity of the leading eigenvalue to baseflow modifications, $\nabla_{\mathbf{U}} \sigma$, as defined in \eqref{eq:sens_base1}-\eqref{eq:sens_base3}. Figure \ref{fig:sens_baseflow} shows the (a) growth rate $\nabla_{\mathbf{U}} \lambda$ and (b) frequency $\nabla_{\mathbf{U}} \omega$ sensitivities at $m=2$ and $k=9.8$ for 0.92 BEP using $\nu_{t,k-\varepsilon}$. Far from the axis ($r>0.5$), the sensitivities decay rapidly to zero as the amplitudes of both direct and adjoint modes diminish, indicating that the instabilities are governed by the inner-core dynamics. The highest magnitudes are observed very close to the axis ($r \sim 0.1$), except for the frequency sensitivity in the axial direction, which peaks at $r \sim 0.2$. Suppose a positive arbitrary baseflow perturbation $\delta \mathbf{U}$ is applied: positive magnitudes in figure \ref{fig:sens_baseflow}(a) correspond to destabilization ($\delta \lambda > 0$), while negative magnitudes indicate stabilization ($\delta \lambda < 0$). Positive perturbations of the radial and azimuthal baseflow components destabilize the flow near the axis but become stabilizing further out. In contrast, positive variations of the axial component $U_z$ are strongly stabilizing near the axis but become slightly destabilizing beyond $r \sim 0.2$. This behavior can be physically linked to the typical vortex rope formation in draft tubes: as the flow rate decreases (moving away from the BEP), the axial velocity decreases, swirl intensity increases, and vortex rope formation is observed \citep{seifi2024linear}. At 0.92 BEP, increasing the axial baseflow and decreasing the azimuthal velocity (swirl intensity) shifts the operating condition closer to the BEP (figure \ref{fig:baseflow}(a)), thereby stabilizing the flow, as in figure \ref{fig:m2_evolution}(a).
\begin{figure}
	\includegraphics[width=\textwidth]{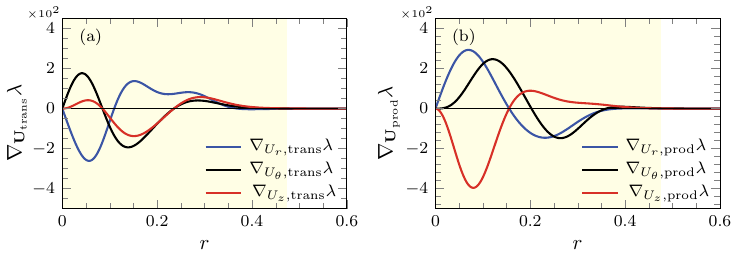}%
	\caption{\label{fig:sens_baseflow_transprod} Sensitivity to baseflow modifications of the growth rate $\nabla_{\tiny\textbf{U}} \lambda$ at $m=2$ and $k=9.8$ for 0.92 BEP using $\nu_{t,k-\varepsilon}$. This sensitivity is decomposed into (a) a sensitivity function to modifications of the transport $\nabla_{\tiny\textbf{U}_{\mathrm{trans}}} \lambda$ and (b) a sensitivity function to modifications of the production $\nabla_{\tiny\textbf{U}_{\mathrm{prod}}} \lambda$ as a function of $r$.}
\end{figure}

Figure \ref{fig:sens_baseflow}(b) shows that the frequency sensitivity is dominated by the azimuthal flow near the core, with a considerable contribution from the axial baseflow around $r \sim 0.2$. As the flow approaches BEP (with decreased swirl intensity and increased axial velocity), negative azimuthal perturbations tend to increase the frequency, whereas axial perturbations tend to decrease it. Since the frequency is more sensitive to azimuthal perturbations, this results to a net increase in frequency along the negative direction, as observed in figure \ref{fig:m2_evolution}(b).

In a simplified sense, the growth rate is most sensitive to perturbations of the axial baseflow, whereas the frequency is primarily influenced by perturbations of the azimuthal baseflow. Physically, this indicates that changes in the axial velocity near the core have the strongest effect on the amplification or decay of instabilities, while variations in the swirl intensity mainly determine the oscillation frequency of the mode. However, the secondary contributions of the other components are not negligible. Similarly, the frequency shows non-negligible sensitivity to axial and radial baseflow perturbations, emphasizing that while a dominant component may govern each aspect of the instability, the interplay between all baseflow components collectively shapes the dynamics of the mode. This analysis is consistent with the wavemaker region in figure \ref{fig:direct_adjoint_mode}(c), confirming that the inner-core region is the primary source of the instability and the most sensitive to baseflow perturbations. At the same time, it provides a more quantitative characterization of how each baseflow component contributes to the growth rate and frequency, enabling a detailed assessment of which modifications would stabilize or destabilize the flow and how they affect the temporal dynamics of the mode.

We can further analyze this local dynamics by going back to the local stability theory (see \cite{huerre1990local}). Locally, a flow is considered absolutely unstable if the production of perturbations dominate the transport of perturbations by the baseflow. \cite{chomaz1988bifurcations} established that this condition is a necessary precursor for the onset of a global instability in a finite spatial domain. Figure \ref{fig:sens_baseflow_transprod} presents the decomposition of the growth rate sensitivity into contributions from (a) transport and (b) production of perturbations. While the radial extents of the two contributions are similar, the sensitivity associated with production clearly dominates, indicating that modifications of the production mechanisms largely govern the overall growth rate sensitivity. This dominance suggests the potential existence of a global instability. Interestingly, for the radial and axial components, the sensitivities to transport and production appear to be reversed. As a result, the radial transport mechanism nearly balances the radial production mechanism, whereas the axial transport mechanism provides a strong stabilizing effect around $r \sim 0.08$. In contrast, the sensitivities associated with azimuthal transport and production are only slightly offset from each other, such that both contributions tend to destabilize the flow near the core before becoming stabilizing further out.

\subsubsection{Sensitivity to turbulent viscosity modifications}
\label{sec:sensitivity_analysis_turb_viscosity}

\begin{figure}
	\includegraphics[width=\textwidth]{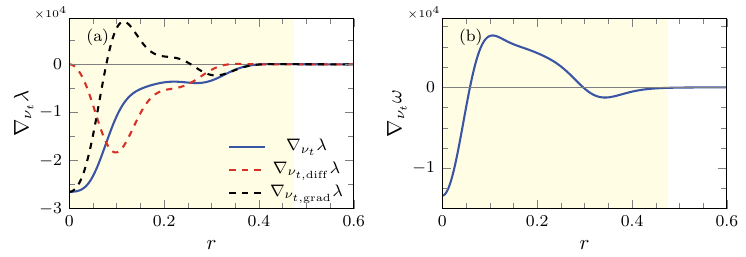}%
	\caption{\label{fig:sens_turbviscosity}  Sensitivity to turbulent viscosity modifications of the leading eigenvalue $\nabla_{\nu_T} \sigma$ at $m=2$ and $k=9.8$ for 0.92 BEP using $\nu_{t,k-\varepsilon}$. (a) Growth rate $\nabla_{\nu_t} \lambda$ and (b) frequency $\nabla_{\nu_t} \omega$ sensitivity to turbulent viscosity modifications as a function of r. The growth rate sensitivity is decomposed into the sensitivity to modifications of the turbulent diffusion $\nabla_{\nu_{t, \mathrm{diff}}} \lambda$ and the sensitivity to modifications of the gradient $\nabla_{\nu_{t, \mathrm{grad}}} \lambda$.}
\end{figure}

As previously seen in figure \ref{fig:specturm}, a dramatic collapse of the growth rate spectrum occurs when we account for turbulent diffusion, regardless of the turbulent model used. This observation calls for a deeper investigation into the role of turbulent viscosity in flow stability. In this study, we explicitly derived the sensitivity of the eigenvalue to modifications in turbulent viscosity. The corresponding results are presented in figure \ref{fig:sens_turbviscosity}, decomposed into (a) growth rate and (b) frequency sensitivities. If we introduce an arbitrary small positive modification of the turbulent viscosity, $\delta \nu_t$, positive magnitudes in figure \ref{fig:sens_turbviscosity}(a) indicate destabilizing effects. The growth rate sensitivity $\nabla_{\nu_t} \lambda$ (blue curve), however, is negative throughout the unstable region, reinforcing our earlier observation that an increase in $\nu_t$ has a stabilizing effect, particularly near the axis. Frequency sensitivity (figure \ref{fig:sens_turbviscosity}(b)) is more dynamic as turbulence reduces oscillation frequency near the axis but increases it in intermediate regions before vanishing further away.

The turbulent viscosity profiles in figure \ref{fig:basevisc}, obtained from the $k-\varepsilon$ relation, vary significantly along the radial direction. As a result, the governing equations include not only turbulent diffusion but also additional terms (see $\mathcal{G}$ in \eqref{eq:G}) associated with the radial gradient of $\nu_t$. This motivates a decomposition of the growth rate sensitivity into these distinct contributions. The broken curves in figure \ref{fig:sens_turbviscosity}(a) represent this decomposition: the sensitivity to modifications of the turbulent diffusion is shown by the red dashed line, while the sensitivity to modifications of the turbulent viscosity gradient is represented by the black dashed line. The turbulent diffusion ($\nabla_{\nu_{t, \mathrm{diff}}} \lambda$) mechanism is mainly stabilizing when viscosity is added, as it damps perturbations across the unstable region. In contrast, the turbulent viscosity gradient ($\nabla_{\nu_{t, \mathrm{grad}}} \lambda$) mechanism is strongly stabilizing near the axis but becomes destabilizing in the intermediate region of instability. This implies that stabilizing the flow near the axis ($r \sim 0$) requires more than pure uniform diffusion alone since $\nabla_{\nu_{t, \mathrm{diff}}} \lambda$ is zero at the axis. Overall, the analysis demonstrates that stabilization is mainly governed by enhanced diffusion, while the gradient contribution, although strongly stabilizing near the axis, can locally induce destabilization in other regions. These results prove that, contrary to conventional intuition, spatial variations in turbulent viscosity can, under certain conditions, promote localized destabilization.

The validation of the sensitivity analyses for both baseflow and turbulent viscosity under arbitrary modifications is provided in Appendix \ref{sec:appen_validation_sensitivity}.

\subsubsection{Sensitivity-based predictions of linear stability}
\label{sec:sensitivity_analysis_predict}

At this stage, we have analyzed the $0.92$ BEP case for $m = 2$ and $k = 9.8$ using $\nu_{t,k-\varepsilon}$, along with the sensitivities of its leading eigenvalue. As a next step, we use these results to predict the growth rate and frequency curves for different baseflow profiles. Specifically, by employing the previously computed sensitivity $\nabla_{U} \sigma$, we can estimate the eigenvalue shift $\delta \sigma$ corresponding to a given baseflow modification $\delta U$ (i.e. the deviation between the current and new baseflow). This procedure is applied across the entire axial wavenumber spectrum to generate predicted growth rate and frequency curves, which can then be compared with the exact linear stability results of the modified baseflow.

\begin{figure}
	\centering
	\includegraphics[width=\textwidth]{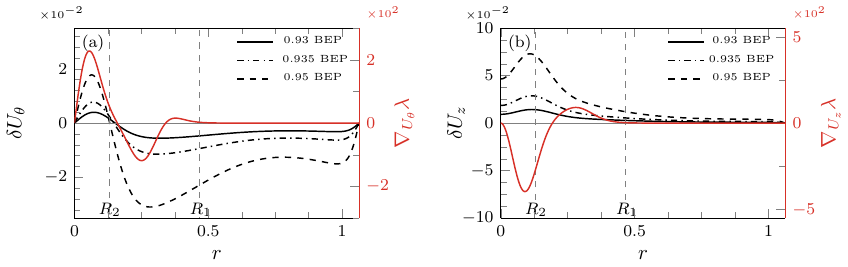}%
	\caption{\label{fig:base} Baseflow modification profiles ($\delta U$) with respect to 0.92 BEP superimposed on the growth rate sensitivity of the leading eigenvalue to baseflow modifications ($\nabla_{U} \lambda$): (a) azimuthal component, (b) axial component.} 
\end{figure}

To remain within the linear regime (small modification), we first apply a fractional step ($\delta$~BEP) from the current baseflow profile. For this purpose, we take $\delta = 0.01$~BEP, yielding a new baseflow profile at 0.93~BEP. In addition, we perform a prediction for 0.935 and 0.95 BEP to examine the behavior of the method under larger modifications. The baseflow parameters, for both velocity and turbulent kinetic energy, for these new conditions are obtained via linear interpolation between 0.92 and 0.98 BEP (see Table \ref{tab:baseflow_params}). Figure~\ref{fig:base} shows the (a) azimuthal and (b) axial velocity perturbation profiles (relative to 0.92~BEP) for the 0.93, 0.935, and 0.95~BEP baseflows, superimposed on the corresponding growth rate sensitivity curves for each velocity component. Only the sensitivity curve for the leading mode at $k = 9.8$ is shown. The predictions, however, are obtained using the sensitivity across the full axial wavenumber spectrum. The complete sensitivity maps are presented in the following section. It should be noted that turbulent viscosity perturbations are also included in the calculation.

\begin{figure}
	\includegraphics[width=\textwidth]{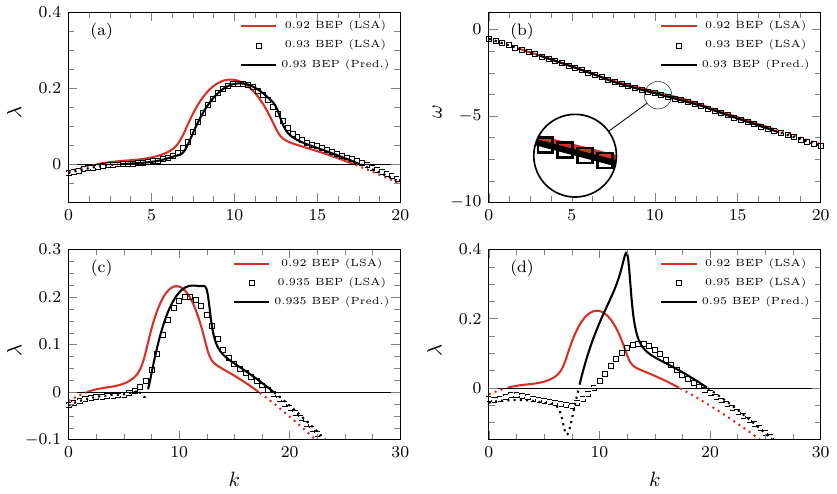}%
	\caption{\label{fig:Predict_idx3_eigen_from_idx2} Predicted versus exact curves of (a) growth rate ($\lambda$) and (b) frequency ($\omega$) for the 0.93 BEP baseflow profile, plotted as functions of the axial wavenumber $k$. The comparisons between predicted and exact growth rates for the (c) 0.935 BEP and (d) 0.95 BEP cases are also shown. Solid lines indicate positive growth rates (and their frequencies), while dotted lines indicate negative growth rates (and their frequencies).}
\end{figure}

Both sensitivities (red curves in figure~\ref{fig:base}) vanish beyond $R_1$, indicating that perturbations introduced at $r>R_1$ have a negligible effect on the leading eigenvalue. Departing from the 0.92 BEP condition leads to progressively larger perturbations, with $\delta U_z$ increasing to as much as 0.08. For the azimuthal component, the perturbations are mostly aligned with the sensitivity direction, resulting in a destabilizing effect ($+\delta \lambda$). In contrast, the axial component exhibits stronger perturbations oriented opposite to the sensitivity direction, producing a stabilizing effect ($-\delta \lambda$). 

Figure~\ref{fig:Predict_idx3_eigen_from_idx2}(a) shows that the growth rate curves for the new baseflow profiles exhibit overall stabilization relative to the previous baseflow while almost no changes are observed for the frequency as shown in figure~\ref{fig:Predict_idx3_eigen_from_idx2}(b). 
The predicted growth rate and frequency curves for the 0.93 BEP case, obtained via sensitivity analysis, show excellent agreement with the exact linear stability results. These observations demonstrate that, once the sensitivity results are known for a given baseflow, one can reliably predict both the magnitude and shape of the growth rate and frequency curves for other baseflow configurations, provided that the perturbations remain within the linear regime.

When the baseflow is modified further as shown in figure~\ref{fig:Predict_idx3_eigen_from_idx2}(c) for 0.935 BEP, the predicted growth rates begin to deviate from the linear stability results, signalling the onset of nonlinear effects. These deviations are even more pronounced in the 0.95 BEP case (figure~\ref{fig:Predict_idx3_eigen_from_idx2}(d)). While the predictions agree well with the linear stability results for $k<6.6$ and $k>14$, there is a noticeable mismatch in the range $k \in [6.6,14]$. This discrepancy is primarily due to the large axial perturbations introduced in the system, which exceed the linear regime. It is therefore necessary to investigate this behavior in detail in the range $k \in [6.6,14]$, which will be addressed in later section \S \ref{sec:sensitivity_analysis_extended}.

\begin{figure}
	\centering
	\includegraphics[width=\textwidth]{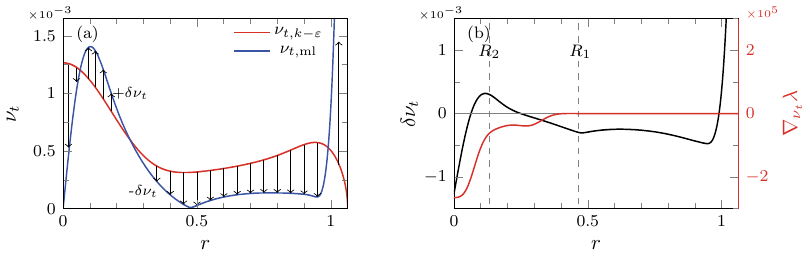}%
	\caption{\label{fig:Nut} (a) Turbulent viscosity profiles ($\nu_{t, k-\varepsilon}$ and $\nu_{t, \mathrm{ml}}$) for 0.92~BEP, emphasizing the differences between the two models and (b) the corresponding perturbation profile ($\delta \nu_t$) superimposed on the growth rate sensitivity of the leading eigenvalue to turbulent viscosity modifications ($\nabla_{\nu_t} \lambda$).}
\end{figure}

We perform a similar analysis for a new turbulent viscosity profile. In this case, we chose to predict the growth rate and frequency curves for the mixing-length turbulent viscosity model, $\nu_{t, \mathrm{ml}}$, using the same baseflow profile for the $m=2$ mode. Figure~\ref{fig:Nut}(a) illustrates how the current $\nu_{t, k-\varepsilon}$ profile can be modified to obtain the mixing-length $\nu_{t, \mathrm{ml}}$ profile, which is equivalent to introducing modification to the turbulent viscosity. The resulting turbulent viscosity perturbation profile is superimposed on the growth rate sensitivity of the leading eigenvalue to turbulent viscosity modifications, as shown in figure~\ref{fig:Nut}(b). Similar to the baseflow sensitivities, the turbulent viscosity sensitivity for the leading growth rate vanishes beyond $R_1$, indicating that perturbations introduced beyond $R_1$ will not affect the growth rate. Within the unstable region, the perturbations are initially aligned with the sensitivity direction, producing a destabilizing effect ($+\delta \lambda$), but gradually reverse orientation, becoming opposite to the sensitivity direction and producing a stabilizing effect ($-\delta \lambda$).

\begin{figure}
	\includegraphics[width=\textwidth]{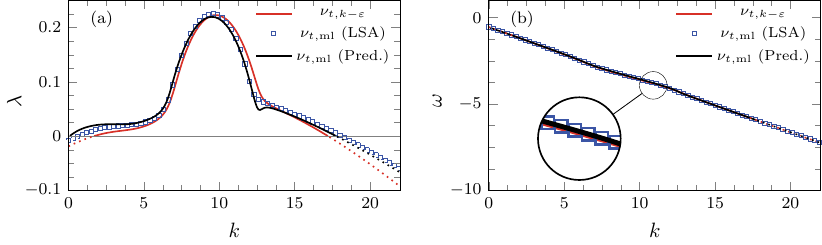}%
	\caption{\label{fig:Predict_NUt_ave_eigen_from_NUt_r} Predicted versus exact (a) growth rate ($\lambda$) and (b) frequency ($\omega$) curves for the mixing length turbulent viscosity profile ($\nu_{t, \mathrm{ml}}$) as a function of axial wavenumber $k$. Solid lines indicate positive growth rates (and their frequencies), while dotted lines indicate negative growth rates (and their frequencies).}
\end{figure}

Figure~\ref{fig:Predict_NUt_ave_eigen_from_NUt_r} compares the predicted growth rate and frequency curves with the results from linear stability analysis for the $\nu_{t, \mathrm{ml}}$ profile. The predicted and exact curves are nearly identical. This observation is particularly interesting because it provides a clear explanation for the similarity of the growth rate spectra in figure~\ref{fig:specturm} between rows~(c) and~(d), which correspond to the $\nu_{t, \mathrm{ml}}$ and $\nu_{t, k-\varepsilon}$ profiles, respectively. At first, this similarity was not obvious, given the substantial differences between the turbulent viscosity profiles (see figure~\ref{fig:basevisc}). The present sensitivity analysis shows that, despite differences between the profiles, the destabilizing and stabilizing contributions within the unstable region balance when integrated over the radial domain. As a result, switching from $\nu_{t,k-\varepsilon}$ to $\nu_{t,\mathrm{ml}}$ produces only minor changes in the growth rate, and the overall growth rate characteristics remain essentially unchanged.

\subsection{Spectral sensitivity distribution across the axial wavenumber space}
\label{sec:sensitivity_analysis_extended}

\begin{figure}
	\includegraphics[width=\textwidth]{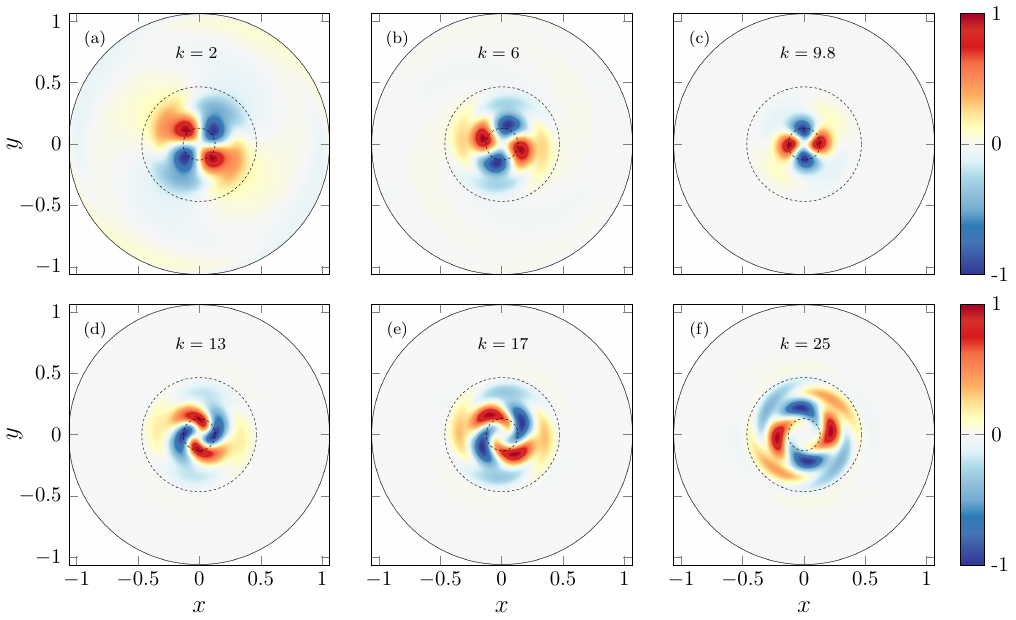}%
	\caption{\label{fig:vorticities_m2} Axial vorticities for different axial wavenumbers ((a) $k=2$, (b) $k=6$, (c) $k=9.8$, (d) $k=13$, (e) $k=17$, (f) $k=25$) for $m=2$. Black dashed circles correspond to $R_2 = 0.13$ and $R_1 = 0.47$ for $0.92\,$ BEP. The vorticity magnitude is  normalized by the absolute maximum value in each case.}
\end{figure}

In \S~\ref{sec:sensitivity_analysis}, we found that, for the most unstable mode, the instabilities are localized near the axis and vanish beyond $R_1$, making them most sensitive to perturbations in the core region. In this section, we examine how this radial behavior evolves across the axial wavenumber $k$.

Figure \ref{fig:vorticities_m2} shows the axial vorticity field for the $m=2$ mode at different axial wavenumbers. For small $k$ ((a) $k=2$, (b) $k=6$, (c) $k=9.8$), the vorticity increasingly concentrates toward the core as $k$ increases, reaching its most compact configuration at $k=9.8$, where the maximum growth rate is observed. As $k$ increases further ((d) $k=13$, (e) $k=17$, (f) $k=25$), the structures become more spiral-like and extend radially outward, gradually forming a hollow core at sufficiently large $k$. This evolution helps explain the large mismatch observed in the growth rate prediction in figure~\ref{fig:Predict_idx3_eigen_from_idx2}(d), particularly for $k \in [6.6,14]$. Most of the baseflow modifications in figure~\ref{fig:base} are localized within the inner core region ($r < R_2$). From the vorticity evolution, we now see that, over this same range of $k$, the instabilities are also concentrated around the core. The combination of relatively large perturbations and the high sensitivity of this region therefore leads to a significant breakdown of the linear prediction. It is also interesting that the vorticity field at $k=9.8$ (figure~\ref{fig:vorticities_m2}(c)), exhibits a less spiral-like and more compact structure, suggesting that this configuration is optimal for extracting the maximum possible perturbation energy from the baseflow as it is associated with the maximum growth rate.

\begin{figure}
	\includegraphics[width=\textwidth]{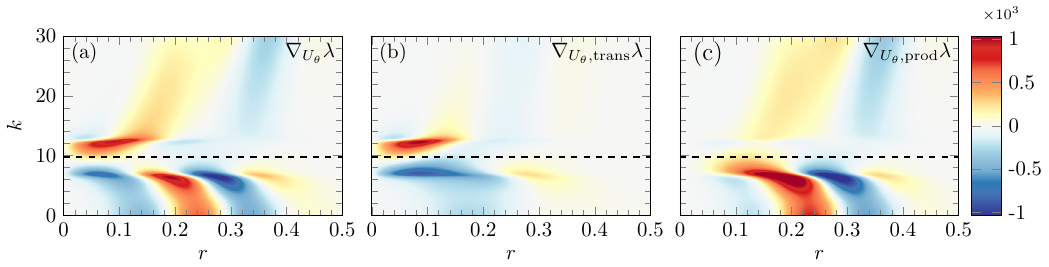}%
	\caption{\label{fig:sens_map_Ut} (a) Growth rate sensitivity to azimuthal baseflow velocity modifications $\nabla_{U_\theta} \lambda$, decomposed into sensitivity to modifications of the (b) transport $\nabla_{U_\theta, \mathrm{trans}} \lambda$ and (c) production $\nabla_{U_\theta, \mathrm{prod}} \lambda$ for $m = 2$ as functions of $k$ and $r$. The dashed line indicates the most amplified axial wavenumber.}
\end{figure}

We now examine in more detail the spectral distribution of the growth rate sensitivity to azimuthal baseflow modifications, $\nabla_{U_\theta}\lambda$, as functions of radius and wavenumber $k$ (figure~\ref{fig:sens_map_Ut}(a)). The most sensitive regions are clearly confined to $r<R_1=0.47$. The location of the axial wavenumber ($k=9.8$) corresponding to the maximum growth rate is indicated by the dashed line, which appears to separate two distinct patterns in the sensitivity field. For small $k$ (below the dashed line), the sensitivity exhibits an alternating radial structure, beginning with a stabilizing region in the core, followed by a destabilizing region, with this alternation occurring twice. For larger $k$ (above the dashed line), the inner core is mostly destabilizing, while the outer region becomes stabilizing. It is also clear that the highest sensitivity occurs for $k \in [6.6,14]$, which supports our earlier interpretation of why the linear prediction breaks down for the 0.95~BEP case in this range.
Figures~\ref{fig:sens_map_Ut}(b-c) presents the decomposition of the growth rate sensitivity into azimuthal transport and production contributions. The transport term is seen to dominate the sensitivity within the core region, while the production term strongly governs the remainder of the sensitive domain farther away from the axis. This interplay between transport and production has also been observed locally at $k=9.8$ in figure~\ref{fig:sens_baseflow_transprod}. Therefore, depending on the radial location, azimuthal baseflow modifications can either act through transport mechanisms in the inner core or through production mechanisms in the outer region, leading to either a stabilizing or destabilizing effect on the leading eigenvalue.

\begin{figure}
	\includegraphics[width=\textwidth]{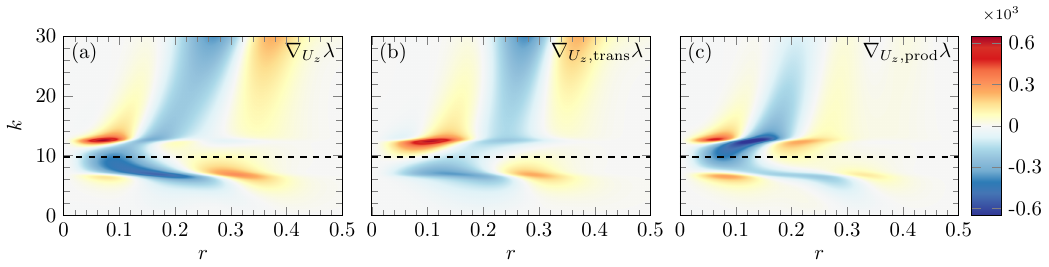}%
	\caption{\label{fig:sens_map_Uz} Same as in figure \ref{fig:sens_map_Ut} but for $U_z$.}
\end{figure}

The sensitivity map for axial baseflow modifications $\nabla_{U_z}\lambda$ is shown in figure~\ref{fig:sens_map_Uz}(a). In the vicinity of $k=9.8$, the inner core is primarily stabilizing, while regions further from the axis are weakly destabilizing. This supports the analysis in the previous section: introducing a positive axial perturbation at 0.92~BEP stabilizes the system, since this effectively increases the flow rate. Away from $k=9.8$, the opposite trend is observed: the core becomes weakly destabilizing, followed by a stabilizing region and then a return to destabilization further outward. Similar to the azimuthal baseflow sensitivity, the highest sensitivity occurs within the range $k \in [6.6,14]$.
Figures~\ref{fig:sens_map_Uz}(b-c) show the respective roles of the axial transport and production contributions to the growth rate sensitivity. Unlike the sensitivity to swirl velocity modifications, which shows a clear separation between production and transport effects across radial locations, these roles are less distinctly separated here. In this case, the production term dominates in the vicinity of the maximum growth rate, whereas the transport term is slightly dominant in other regions. In particular, around $k=9.8$, where the largest growth rates are observed, the production mechanism is more dominant than the transport contribution.

\begin{figure}
	\includegraphics[width=\textwidth]{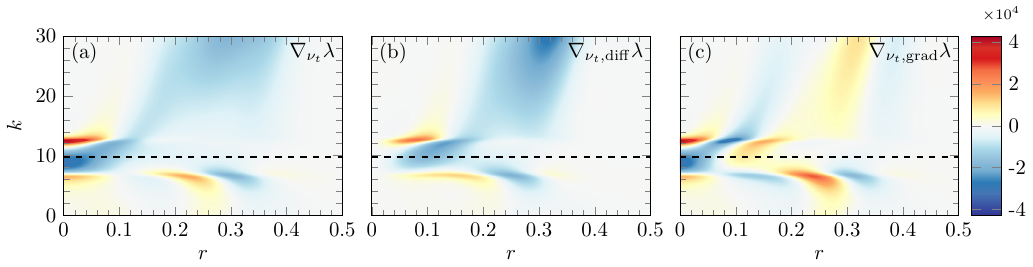}
	\caption{\label{fig:sens_map_NUt} (a) Growth rate sensitivity to turbulent viscosity modifications $\nabla_{\nu_t} \lambda$, decomposed into sensitivity to modifications of the (b) diffusion $\nabla_{\nu_t, \mathrm{diff}} \lambda$ and (c) gradient $\nabla_{\nu_t, \mathrm{grad}} \lambda$ for $m = 2$ as functions of $k$ and $r$. The dashed line indicates the axial wavenumber corresponding to the location of the maximum growth rate.}
\end{figure}

Finally, we investigate the spectral distribution of the growth rate sensitivity to turbulent viscosity modifications, $\nabla_{\nu_t} \lambda$, as shown in figure \ref{fig:sens_map_NUt}(a). The major portion of the figure is bluish indicating negative value. Thus, the growth rate sensitivity to positive turbulent viscosity modifications is mostly stabilizing. Figures \ref{fig:sens_map_NUt}(b-c) shows the decomposition of this sensitivity to diffusion and gradient mechanisms. The diffusion contribution largely contributes to the overall stabilizing effect of turbulent viscosity. However, near the axis, the gradient mechanism is mainly responsible for this stabilizing effect. In certain regions however, it induces strong destabilization as partly shown in \S \ref{sec:sens_turb_viscosity}. This behavior appears to be related to the steep downward slope (at $r \in [0.1,0.2]$) of the $\nu_t$ profile (figure \ref{fig:basevisc}), thereby rapidly decreasing $\nu_t$ and resulting to a local destabilizing effect across this region. Therefore, if one wants to stabilize the flow very near the axis, it is important to take into account the spatial variation of turbulent viscosity.

\subsection{Stability analysis with a reduced turbulent length scale $\ell = 0.01$}
\label{sec:lower_turb_length}

In the previous sections, the turbulent length scale $\ell = 0.02$ was employed in the $k-\varepsilon$ relation, as it provides the best agreement with experimental measurements downstream in the conical diffuser as reported by \cite{Susan10}. However, several studies have reported smaller values. In particular, \cite{mauri2004werle} suggested $\ell = 0.01$ based on laser Doppler velocimetry measurements and estimates of the turbulent eddy size between runner blades. Other authors \citep{benim1990finite} also reported the same values. In order to compare the present results to the choice of turbulent length scale, we briefly discuss computations performed using $\ell = 0.01$.

\begin{figure}
	\centering
	\begin{subfigure}{\textwidth}
		\includegraphics[width=\textwidth]{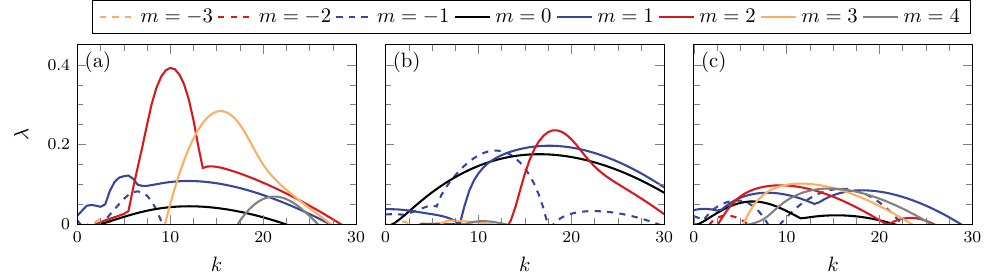}
	\end{subfigure}
	\caption{\label{fig:turb_0pt01_spectrum} Growth rate spectra $\lambda$ as functions of the axial wavenumber $k$ and azimuthal wavenumber $m$ for the operating conditions: (a) $0.92$ BEP, (b) BEP, and (c) $1.06$ BEP using $\nu_{t,k-\varepsilon}$ with $\ell = 0.01$.}
\end{figure}

Figure \ref{fig:turb_0pt01_spectrum} presents the growth rate spectra as functions of the axial wavenumber for the three operating conditions, computed using the $\nu_{t,k-\varepsilon}$ profile. The overall structure of the spectra is qualitatively similar to that observed in our earlier results (figures \ref{fig:specturm}(1d-3d)). However, the growth rates are systematically higher, in some cases nearly doubled. In addition, several higher azimuthal modes become unstable, with unstable modes now observable up to $m = 4$ for all three baseflows. The reduced turbulent length scale leads to a lower effective turbulent viscosity, thereby weakening viscous damping and allowing a broader range of instability to develop with enhanced growth.

\begin{figure}
	\includegraphics[width=\textwidth]{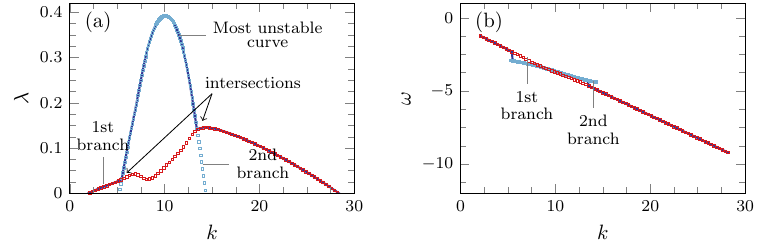}%
	\caption{\label{fig:mt_2_eigenvalues} (a) Growth rate and (b) frequency curves for the baseflow profile ($0.92\,\mathrm{BEP}$) using $\nu_{t,k-\varepsilon}$ with $\ell = 0.01$ for azimuthal wavenumber $m=2$ as a function of $k$. The solid line shows the most unstable growth rate and the corresponding frequency. The symbols are for the first two most unstable modes.}
\end{figure}
\begin{figure}
	\includegraphics[width=\textwidth]{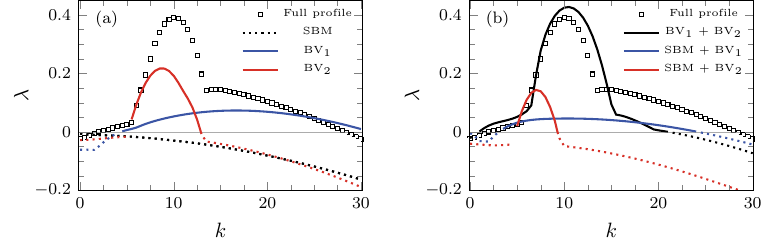}%
	\caption{\label{fig:vortex_combination} Growth rate spectra of the decomposed baseflow profile at $m=2$ ((a) solid-body motion and individual Batchelor vortices, (b) their pairwise combinations). SBM: Solid-body motion, $\mathrm{BV}_1$: Batchelor vortex 1, $\mathrm{BV}_2$: Batchelor vortex 2. Solid lines indicate positive growth rates (and their frequencies), while dotted lines indicate negative growth rates (and their frequencies).}
\end{figure}

Let us examine in more detail the results obtained for the 0.92 BEP baseflow profile using the $\nu_{t,k-\varepsilon}$ profile with a turbulent length scale of $0.01$, as shown in figure \ref{fig:mt_2_eigenvalues}. The maximum growth rate curve is composed of two distinct and intersecting branches over the range of axial wavenumbers considered. This feature was not observed in the $\ell=0.02$ case, indicating that the existence of multiple instability branches is sensitive to the level of turbulent diffusion included in the model. 
The presence of distinct curves governing the stability in different intervals of $k$ suggests the coexistence of multiple competing instability mechanisms.
 As discussed in \S \ref{sec:baseflow}, the baseflow is composed of three distinct vortical structures, each of which may contribute differently to the overall instability dynamics. It is therefore interesting to investigate whether these branches can be directly associated with the individual vortices constituting the baseflow, and to identify the role played by each structure in driving the observed instabilities.

Figure~\ref{fig:vortex_combination} shows the growth rate spectra of the decomposed baseflow profile. Here, $\mathrm{BV}_1$ and $\mathrm{BV}_2$ denote the two counter-rotating Batchelor vortices associated with ${R}_1$ and ${R}_2$, respectively, where ${R}_1 > {R}_2$, and SBM refers to the solid-body motion. In figure~\ref{fig:vortex_combination}(a), the solid-body motion alone does not sustain any unstable modes but it contributes on the frequency curve of the full profile (not shown). The contribution of $\mathrm{BV}_2$ is clearly illustrated by the bell-shape structure observed at low axial wavenumbers. Its frequency is almost constant (not shown), except at the region where it is unstable. In contrast, $\mathrm{BV}_1$ generates a smooth growth rate curve across a broader range but without the bell-shape feature at low $k$ while maintaining a constant frequency (not shown). This indicates that $\mathrm{BV}_2$ is primarily responsible for the bell-shaped small $k$ instability, whereas $\mathrm{BV}_1$ acts to amplify the instability and extend it towards higher axial wavenumbers. It should be noted that neither of the individual vortices displays the two-branch structure observed in the full baseflow, indicating that the emergence of multiple branches is a consequence of the interaction between the three components of the baseflow, rather than a property of any single vortex in isolation.

Figure~\ref{fig:vortex_combination}(b) further supports this interpretation: the combined effect of $\mathrm{BV}_1$ and $\mathrm{BV}_2$ already reproduces most features of the full profile, but the maximum growth rate is amplified while the growth rates are damped at higher axial wavenumber $k>15$. The addition of a solid-body motion to the individual Batchelor vortices damps their growth rate curves, especially at higher axial wavenumber. In terms of frequency, the resulting curve only follows the full baseflow profile frequency curve when SBM is included; otherwise, it remains almost constant  (not shown).

\subsection{Baseflow profiles and instability parameters across the flow rate spectrum}
\label{sec:baseflow_discharge_coeff}
As a final analysis, we examine the baseflow features across the flow rate domain using inviscid theories. The baseflow parameters are represented by fitted curves as functions of the flow rate with respect to BEP point and are given in Appendix \ref{sec:curve_fit}.  Figure \ref{fig:baseflow_dc1} visualizes the continuous variation of the baseflow ($U_\theta$ and $U_z$) and the corresponding turbulent viscosity ($\nu_{t,k-\varepsilon}$) profiles across the flow rate and radius with color indicating intensity. In figure \ref{fig:baseflow_dc1}(a), we can observe that below the BEP (indicated by the horizontal solid line), the positive swirl intensifies, increasing the centrifugal forces within the core. This tends to destabilize the flow locally by enhancing the amplification of azimuthal disturbances. Above the BEP, the positive swirl weakens while negative swirl intensifies, creating stronger shear layers and regions of adverse swirl gradient, which also promotes instability. The BEP represents a configuration where the net swirl intensity is balanced, minimizing destabilizing swirl gradients. Figure \ref{fig:baseflow_dc1}(b) shows how the axial velocity increases as the flow rate increases. The axial flow acts to advect disturbances downstream, influencing their growth. Below the BEP, the stronger positive swirl can be stabilized by increasing the axial velocity, which transports perturbations more rapidly and reduces their growth time. In contrast, above the BEP, the flow can be stabilized by decreasing the axial velocity, limiting the interaction with the intensified negative swirl. At the BEP, instability is minimized because the axial and azimuthal velocity profiles are optimally balanced. The swirl gradients are moderate, and the axial advection neither under-advects nor over-accelerates disturbances. The complete turbulent viscosity map based on the $k-\varepsilon$ relation is also provided in figure \ref{fig:baseflow_dc1}(c), exhibiting a strong viscous effect near the core at lower flow rate.

\begin{figure}
	\includegraphics[width=\textwidth]{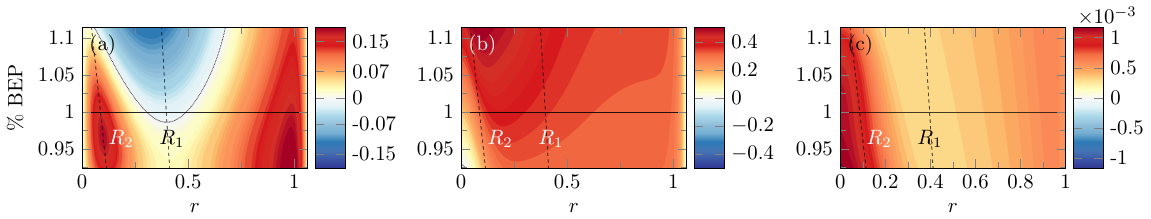}%
	\caption{\label{fig:baseflow_dc1} (a) Azimuthal $U_\theta$, (b) axial $U_z$ and (c) turbulent viscosity $\nu_{t,k-\varepsilon}$ profiles as functions of the flow rate (relative flow rate represented by \% BEP) and $r$. The solid and dashed lines indicate the BEP and vortex core radii ($R_1$, $R_2$), respectively. Dotted line in (a) is the zero-contour curve.
	}
\end{figure}
\begin{figure}
	\centering
	\includegraphics[width=\textwidth]{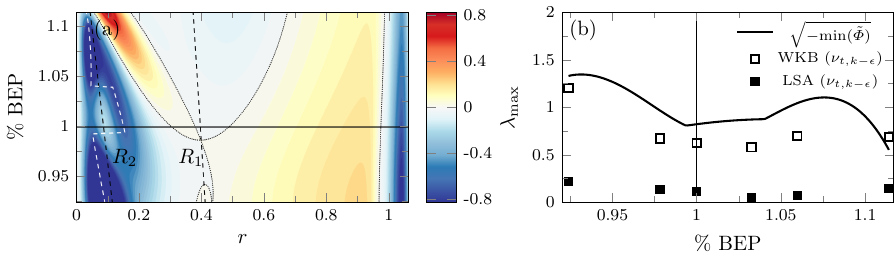}%
	\caption{\label{fig:pred_growthrate_phi_tilde}(a) Leibovich \& Stewartson instability condition $\tilde{\varPhi}$  as functions of relative flow rate \% BEP and $r$. The solid line indicates the BEP, the dashed lines denote the radii $R_1$ and $R_2$, the white dashed line marks $\min(\tilde{\Phi})$, and the dotted lines show the zero-contour curve. (b) The predicted maximum growth rate $\lambda_{\mathrm{max}}$ from $\tilde{\varPhi}$, WKB, and LSA as a function of \% BEP. }
\end{figure}

To further analyze the instability trend, the sufficient condition for instability $\tilde{\varPhi} < 0$ \eqref{eq:condition}, is illustrated in figure \ref{fig:pred_growthrate_phi_tilde}(a) across the entire baseflow condition range. The blue regions indicate where $\tilde{\varPhi}<0$, and darker shading corresponds to a higher maximum growth rate. Below the BEP, the region where $\tilde{\Phi}<0$ is mainly localized near the core (approximately $r<R_2$) and progressively shrinks as the flow rate increases toward the BEP. Around the BEP, two local minima are observed: a new minimum appears near $r\sim0.2$, and a weakly negative $\tilde{\Phi}$ develops in the mid-radius range ($r\sim0.5$) and expands as the flow rate increases further. At the highest flow rate, the instability criterion becomes squeezed into a narrower radial region.

\begin{figure}
	\includegraphics[width=\textwidth]{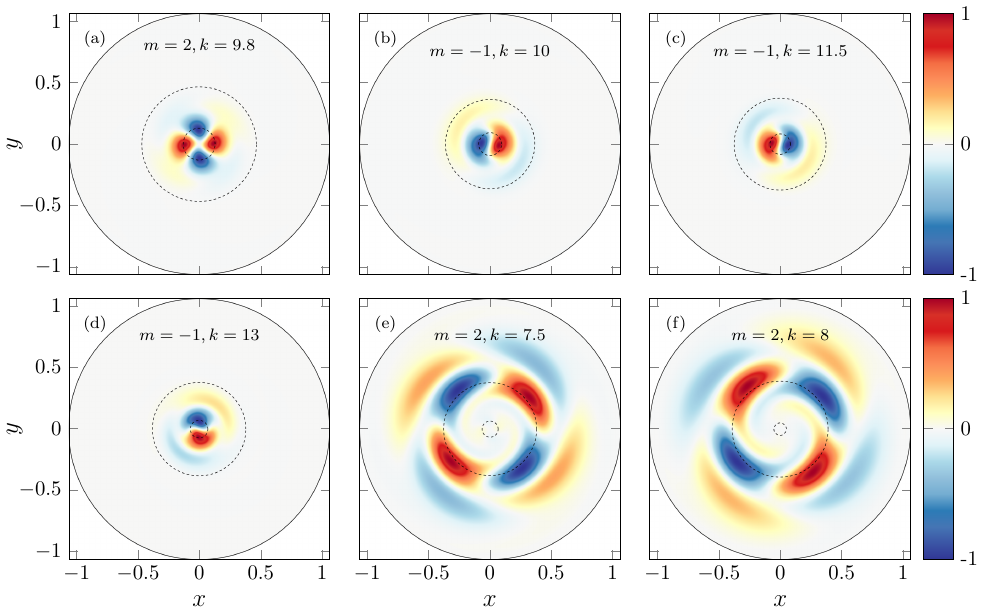}%
	\caption{\label{fig:vorticities_dc} Axial vorticities for different flow rate conditions ((a) 0.92 BEP, (b) 0.98 BEP, (c) 1.00 BEP, (d) 1.03 BEP, (e) 1.06 BEP, (f) 1.11 BEP) for the most unstable mode for each baseflow configuration. Black dashed circles correspond to the vortex cores $R_1$ and $R_2$. The vorticity magnitude is  normalized by the absolute maximum value in each case.}
\end{figure}

From the same figure, $\lambda_{\mathrm{max}}$, defined in \eqref{eq:max_lambda}, is extracted (white dashed lines) and plotted in figure \ref{fig:pred_growthrate_phi_tilde}(b) as a function of relative flow rate (\% BEP), in comparison with WKB and LSA results. Although it is an inviscid first approximation, a local minimum $\lambda_\mathrm{max}$ is found around the BEP. Below BEP, a steep $\lambda_\mathrm{max}$ gradient is observed while a slowly increasing slope is first observed above BEP before it increases rapidly. The $\lambda_\mathrm{max}$ decrease after 1.08 BEP is linked to the instability region where $\tilde{\varPhi} < 0 $ is narrowed (see figure \ref{fig:pred_growthrate_phi_tilde}(a)). The maximum growth rates from WKB (for $\kappa \gg 1$) and linear stability analysis for the operating points are also shown in the plot. 
Although the WKB analysis provides limited improvement over the LSA results, it better captures the qualitative growth rate behaviour than the simplest inviscid criterion based on $\min(\tilde{\Phi})$.

Finally, figure \ref{fig:vorticities_dc} presents the axial vorticity structures of the most unstable modes at different flow rates. At the BEP, as well as slightly below (0.98 BEP) and above (1.03 BEP) it, the most unstable mode corresponds to the $m=-1$ azimuthal wavenumber. At lower flow rate (0.92 BEP), the $m=2$ mode becomes the most unstable. This is similarly observed at higher flow rates (1.06 BEP and 1.11 BEP) but the instabilities are now concentrated outside the core region as the instability criteria shift to larger radii, making a hollow core less sensitive to perturbations. Overall, the most unstable modes are consistently observed within the range $k \approx 10 \pm 3$.

In this section, we demonstrated that just by knowing the distribution of the baseflow and the corresponding instability parameter, one can (1) infer the location of the BEP and the corresponding instability characteristics through analysis of the flow evolution and (2) quickly estimate the maximum growth rates at the inviscid limit across the flow rate spectrum, all without performing a full linear stability analysis. However, when the flow is inherently turbulent, a full linear stability analysis with a turbulent eddy viscosity is then necessary to correctly describe the stability of the system.

\section{Conclusion}
\label{sec:conclusion}

In this study, a linear stability and sensitivity analyses were conducted for a swirling jet for the flow in a Francis turbine draft tube. This work is motivated by the need to better understand draft tube flow dynamics which is linked to the overall efficiency of the hydropower plant. As a starting point, we used the local measurements at the draft tube inlet reported by \cite{Susan06} to perform a stability analysis, addressing whether the measured turbulent mean flow can provide more information on the unsteady dynamics through a linear stability analysis.

We first selected three mean flow profiles corresponding to operating points below, at, and above the best efficiency point (BEP) flow rate, and investigated the effect of incorporating turbulent viscosity into the linear stability analysis. It has been shown that the presence of an eddy viscosity field that models the turbulent interactions greatly reduces the growth rates predicted by nearly inviscid analysis. The range of unstable azimuthal wavenumbers is also reduced to $m \in [-1, 2]$ which better links to the experimental evidence. Three approaches of incorporating the eddy viscosity $\nu_t$ were compared (constant $\overline{\nu_t}$, mixing length model computed from the baseflow, $\nu_{t,\mathrm{ml}}$, and experimentally measured $\nu_{t,k-\varepsilon}$ eddy viscosities). Qualitatively similar result were obtained with all three approaches, with very close agreement of stability properties between $\nu_{t,\mathrm{ml}}$ and $\nu_{t,k-\varepsilon}$. The partial load regime (0.92 BEP) was identified as the most unstable configuration for all three $\nu_t$ profiles, illustrating that a simple mixing length model can capture the essential stability properties of the swirling flow in turbulent regimes. Similar conclusion was draw in previous studies on wind turbine wakes \citep{Viola14}. 
  
To further assess the influence of turbulent eddy viscosity and the changing operation points on the stability spectrum, we performed the adjoint-based sensitivity analysis. The sensitivity analysis to baseflow modifications localizes the regions of the flow most susceptible to instability and indicates that modifications to the axial velocity component strongly influence the growth rate, whereas changes to the azimuthal component primarily affect the frequency. Moreover, the dominant contribution of the production mechanism over the transport mechanism could indicate the existence of a global instability. The sensitivity to turbulent viscosity modifications is derived and showed that accounting for spatial variations of $\nu_t$ improves the accuracy of the range of azimuthal wavenumbers that lead to a linear instability. The analysis also showed that, near the axis where the instabilities are concentrated, adding turbulent diffusion alone without considering spatial gradient of turbulent viscosity profile is insufficient to stabilize the eigenmodes near the core. In addition, spatial variations in turbulent viscosity can, in certain regions, promote destabilization.

Practical discussions are also presented on how the results of the sensitivity analysis can be employed to predict the growth rate and frequency curves of a modified baseflow. It has been shown that the change of stability properties of the baseflow upon a small perturbation can be reliably predicted by the gradient computed in the adjoint sensitivity analysis. This prediction holds as long as the baseflow increment corresponds to up $1\%-2\%$ BEP increment. It should be noted that increasing the operating point in terms of \% BEP inherently introduces perturbations in the baseflow fields $U_{\theta}$, $U_{z}$, and in the turbulent viscosity $\nu_{t}$. The accuracy of this approach is shown to be closely linked to the locations of the instabilities along the axial wavenumber space. The vorticity distributions further qualitatively show that, as the axial wavenumber approaches the location of the maximum growth rate, the instabilities become increasingly concentrated within the inner vortex core. As the wavenumber increases beyond this value, the instabilities gradually shift radially outward, ultimately forming a hollow core surrounded by the unstable outer layers. This progression illustrates how the spatial structure of the instability varies with axial wavenumber, giving emphasis on radial regions most susceptible to perturbations.

Further discussions of the WKB analysis and inviscid theories indicated that these approaches can still provide a general understanding of the flow instability. The WKB analysis successfully predicted the viscous cut-off beyond which the instability ceases to exist, the axial wavenumber of the most unstable mode, and the overall decay at high axial wavenumbers, but overestimates the growth rate magnitudes for small $k$, which is attributed to the breakdown of its underlying assumptions in that regime. These analyses also emphasized that turbulence viscosity greatly influences flow stability and that inviscid theories fail to capture the strong damping effects of viscosity, reaffirming the necessity of incorporating eddy viscosity effects into linear stability.

Analysis of the baseflow profiles across the flow rates is also performed using predictions from inviscid theory. One can qualitatively infer the maximum growth rate curve and identify trends such as the location of the BEP and regions of increased instability, without performing a full linear stability analysis. The inclusion of turbulent viscosity, however, alters these trends, effectively damping growth rates and modifying the spatial structure of the unstable modes.

In this study, we deliberately restrict the analysis to the measured local radial profiles. This keeps the approach extremely simple, as it is essentially one-dimensional (a full stability curve can be obtained within minutes), and it also facilitates comparison with analytical approximations such as WKB analysis and inviscid instability criteria. However, the approach is limited because it does not account for the diverging draft tube geometry and its influence on the flow stability. As a natural extension of this work, a global stability analysis shall be performed. The sensitivity methods developed here can then be applied in the global stability setting to identify the essential elements of eddy viscosity modeling in capturing $m=1$ and $m=2$ vortex rope modes observed experimentally. Ultimately, it would also be of interest to design flow control strategies for stabilization based on the sensitivity information.

\bibliographystyle{jfm}
\bibliography{LSA_turbine_vortex_ref}

\appendix
\section{Validation of the FreeFEM++ code with Chebyshev spectral method}
\label{sec:appen_freefem_validation}

In this section, validations of FreeFEM++ code against the Chebyshev pseudo-spectral collocation method \citep{antkowiak2004transient, Ok15a} is presented for the Batchelor vortex profile \citep{batchelor1964axial}, which is a part of baseflow vortex \eqref{eq:vr}-\eqref{eq:vz}. Its dimensionless velocity profiles are expressed as
\begin{equation}
	\mathrm{U}_r = 0, \quad
	\mathrm{U}_\theta(r) = \frac{q}{r} e^{-r^2}, \quad
	\mathrm{U}_z(r) = e^{-r^2},
\end{equation}
where the swirl parameter $q$ is a non-dimensional measure of the core rotation rate. The flow is inviscidly unstable for $0 < q \leq 1.5$, with the strongest instability observed at $q = 0.87$ \citep{Billant_Gallaire_2013}. For validation, we select $q = 0.8$ at Reynolds number ${Re} = 667$, following \cite{delbende1998absolute}.

\begin{figure}
	\includegraphics[width=\textwidth]{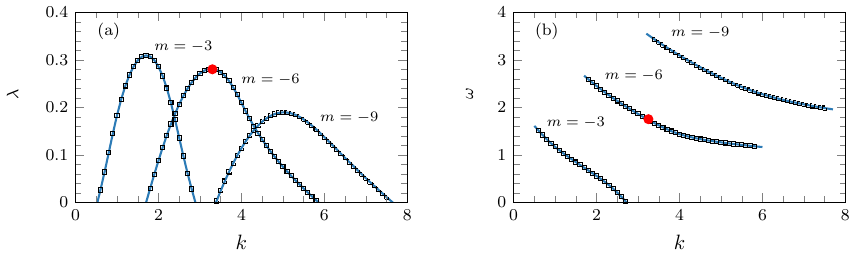}%
	\caption{\label{fig:validation_Batchelor_eigenvalues} (a) Growth rate $\lambda$ and (b) frequency $\omega$ spectrum for azimuthal wavenumbers $m=-3,\ -6,$ and $-9$ for the Batchelor vortex $(q=0.8)$ at ${Re}=667$. Results obtained using the Chebyshev spectral method are shown as square symbols, while those from the current FreeFEM++ implementation are represented by solid lines. The red dot indicates the leading eigenvalue for $m=-6$ reported by \cite{delbende1998absolute}.}
\end{figure}

Figure \ref{fig:validation_Batchelor_eigenvalues} shows the comparison of the growth rate and frequency spectrum as a function of axial wavenumber $k$ obtained using the Chebyshev spectral method and the current FreeFEM++ implementation for different azimuthal wavenumbers $m=[-3,-6,-9]$. Excellent agreement is observed for the eigenvalue spectrum results, validating the accuracy of our current numerical approach. 

\begin{figure}
	\includegraphics[width=\textwidth]{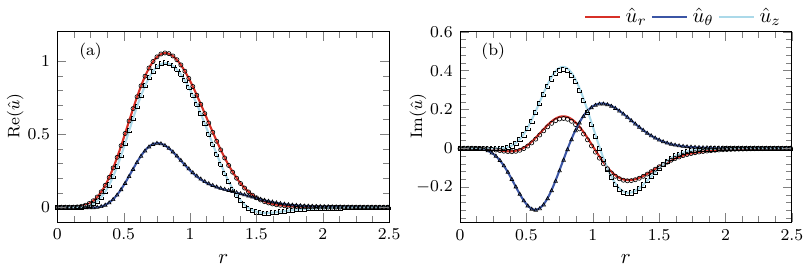}%
	\caption{\label{fig:validation_Batchelor_with_spectral} Comparison between the eigenvectors obtained using the Chebyshev spectral method (solid lines) and the FreeFEM++ code (markers). Shown are (a) the real part $\mathrm{Re}(\hat{u})$ and (b) the imaginary part $\mathrm{Im}(\hat{u})$ of the eigenvectors for the Batchelor vortex $(q = 0.8)$, for $m = -6$ and $k = 3.22$ at ${Re}=667$.}
\end{figure}

A comparison of the eigensolutions computed with FreeFEM++ and the Chebyshev code corresponding to the most amplified wavenumber $k = 3.22$ for $m = -6$, with an associated eigenvalue $\sigma=0.28 + 1.75\, \mathrm{i}$, is shown in figure~\ref{fig:validation_Batchelor_with_spectral}. This eigenvalue is also reported in \cite{delbende1998absolute}, marked with a red dot in figure \ref{fig:validation_Batchelor_eigenvalues}. 
Since the eigenvectors are defined up to a multiplication with a complex constant, the eigenvector is normalised to $\langle \hat{\bm u},\hat{\bm u}\rangle=1$ and its phase is fixed so that $\int\left(\hat{u}_r + \hat{u}_\theta + \hat{u}_z\right)\,r dr$ is real. Excellent agreement between the FreeFEM++ implementation, the Chebyshev spectral method, and results of \cite{delbende1998absolute} validate the methodology implemented in the current work.

\begin{figure}
	\includegraphics[width=\textwidth]{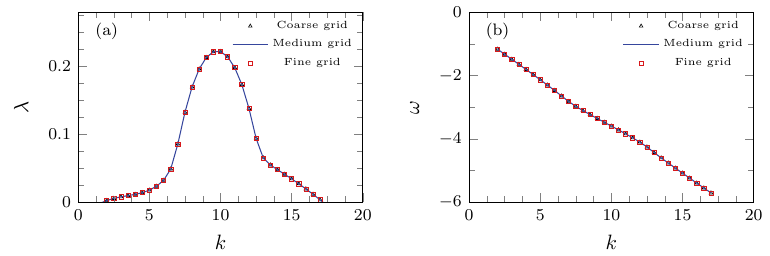}%
	\caption{\label{fig:grid_study} (a) Growth rate and (b) frequency curves from the grid sensitivity study using coarse ($500$ elements), medium ($2,000$ elements), and fine ($10,000$ elements) grids for $0.92\, \mathrm{BEP}$ and $m = 2$ (configuration as in figure \ref{fig:specturm}(1d)). Results superimpose independent of the grid resolution.
	}
\end{figure}

\section{Grid convergence study}
\label{sec:appen_grid_study}

A grid convergence study was performed prior to simulating the various cases presented in this paper, to ensure that the results are not affected by domain resolution. Figure~\ref{fig:grid_study} displays the growth rate and frequency curves for $m = 2$ at $0.92\,\mathrm{BEP}$. The eigenvalue spectrum remained consistent across different grid configurations, confirming grid independence. 

The medium grid configuration ($2{,}000$ elements) was selected for most of the computations reported in this work as it did not incur significant computational cost compared to the coarse grid. The relative errors in the computed eigenvalues for the coarse and medium grids, compared to the fine grid, were very small (less than $0.7\%$), indicating that the results are essentially independent of grid resolution.

\section{Validation of the sensitivity analysis}
\label{sec:appen_validation_sensitivity}
In this section, we validate the sensitivity analysis obtained in \S \ref{sec:sensitivity_analysis}. To do so, we introduced a small arbitrary modification to the baseflow and see if we can predict the actual eigenvalue shift. Any perturbation profile could be chosen, here, we adopt a Gaussian function defined as
\begin{equation}
	\delta \alpha = \mathrm{exp}\left(\frac{-\left( r - r_0 \right)^2}{2\gamma^2}\right), 
\end{equation}
where $\delta \alpha $ is any of $ \delta \mathrm{U}_r, \delta \mathrm{U}_\theta, \delta \mathrm{U}_z, \delta \nu_T$, and the Gaussian is centered at $r_0$ and its width is controlled by $\gamma$. The magnitude of the perturbation is scaled by $\epsilon$. The change in the eigenvalue is then given by
\begin{eqnarray}
	\delta \sigma & = & \lim_{\epsilon \rightarrow 0} \underbrace{\sigma (\alpha + \epsilon \delta \alpha) - \sigma (\alpha) }_{\delta \sigma_{\mathrm{actual}}} = \lim_{\epsilon \rightarrow 0} \underbrace{\left\langle \nabla_{\alpha} \sigma, \epsilon \delta \alpha \right\rangle}_{\delta \sigma_{\mathrm{predicted}}}, \\
	\mathrm{error} & = &  \delta \sigma_{\mathrm{actual}} - \delta \sigma_{\mathrm{predicted}} = \mathrm{error}(\delta \lambda) + i\ \mathrm{error}(\delta \omega).
\end{eqnarray}

The actual eigenvalue variation is computed from the full linear stability analysis, while the predicted variation is obtained from the sensitivity formulation. The sign of the growth rate shift ($\delta \lambda$) indicates whether the prediction correctly captures the stabilizing ($-\delta \lambda$) or destabilizing ($+\delta \lambda$) trend shown by the sensitivity maps.

In our validation test, we set $\gamma = 0.01$ to approximate a local perturbation of the field, thereby facilitating the identification of the directionality of the eigenvalue shift around a point location. The perturbation exercise is carried out at three radial locations: $r_0 = 0.1$, $0.2$, and $0.3$. The test points are chosen near the axis since this is the most sensitive region for the leading eigenmode.

\begin{figure}
	\includegraphics[width=\textwidth]{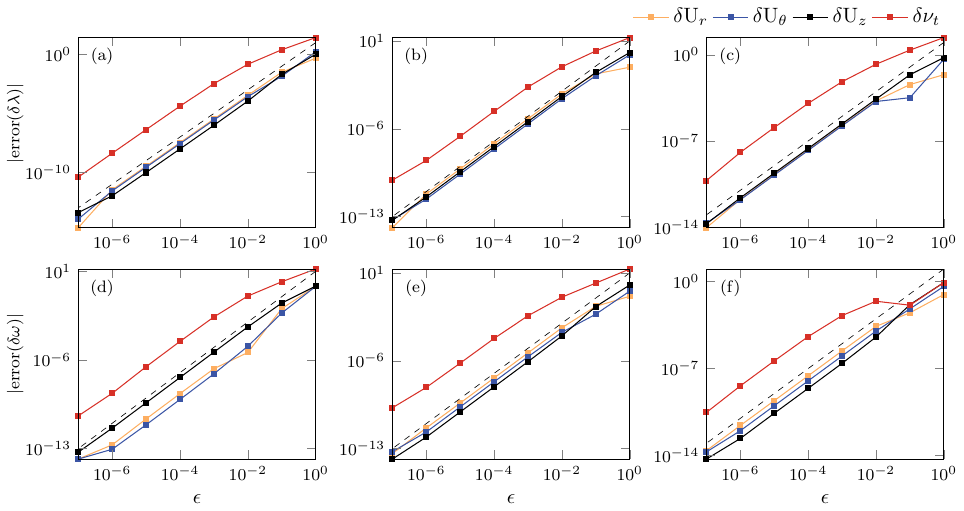}%
	\caption{\label{fig:validation_Sens_baseflow} Convergence plot of the prediction errors for the baseflow perturbation $[\delta \mathrm{U}_r, \delta \mathrm{U}_\theta, \delta \mathrm{U}_z]$ and turbulent viscosity perturbation $(\delta \nu_T)$, as functions of the perturbation magnitude $\epsilon$ for $\gamma = 0.01$ at different radial locations. Panels (a)-(c) show the growth rate error (error($\delta(\lambda)$)) for $r_0 = 0.10$, $0.20$, and $0.30$, respectively, while (d)-(f) show the corresponding frequency error (error($\delta(\omega)$)) at the same locations. The black dashed line indicate the second-order scaling (slope = 2). }
\end{figure}

Figure \ref{fig:validation_Sens_baseflow} shows the convergence plot of the prediction error for the baseflow perturbation $[\delta \mathrm{U}_r, \delta \mathrm{U}_\theta, \delta \mathrm{U}_z]$ and turbulent viscosity perturbation $(\delta \nu_T)$ as a function of the perturbation magnitude $\epsilon$ for $\gamma = 0.01$ at different radial locations. Overall, the figure demonstrates that the prediction of growth rate and frequency variations is accurate and convergent for sufficiently small perturbations, with expected quadratic scaling before numerical errors dominate. The observed second-order convergence (slope = 2) reflects the truncation of the eigenvalue expansion at first order in the sensitivity analysis, with the dominant error arising from the neglected second order terms in the baseflow and turbulent viscosity perturbations. The baseflow velocity components are of order $10^{-1}$, whereas the turbulent viscosity is of order $10^{-3}$, which explains the observed difference in error magnitudes for the same $\epsilon$ values.

\section{Curve fitting of the baseflow parameters}
\label{sec:curve_fit}
The fitted coefficients used in \S \ref{sec:baseflow_discharge_coeff}, following \cite{Susan10}, are summarized below. Each parameter is expressed as a polynomial function of the flow coefficient $\varphi$ and normalized using the reference scales $\Omega_{\mathrm{ref}}$, $U_{\mathrm{ref}}$, and $R_{\mathrm{ref}}$. The flow coefficient $\varphi$ is dimensionless flow rate which links to our relative flow rate defined as \% BEP $=\varphi/\varphi_{\mathrm{BEP}}$ where $\varphi_{\mathrm{BEP}}=0.368$. Therefore, the valid range of $\varphi$ is $\varphi = [0.338, 0.408]$ corresponding \% BEP = [0.92 BEP, 1.11 BEP].
\begin{table}
	\centering
	\begin{tabular}{cl}
		\toprule
		\textbf{Parameter} & \textbf{Fitting formula} \\
		\midrule
		$\Omega_0$ & $(-16.37\,\varphi + 35.38)/\Omega_{0}$ \\
		$\Omega_1$ & $(-427.4\,\varphi + 75.82)/\Omega_{0}$ \\
		$\Omega_2$ & $(-1.489\times10^{5}\varphi^{2} + 1.098\times10^{5}\varphi - 1.985\times10^{4})/\Omega_{0}$ \\
		\midrule
		$W_0$ & $(8.746\,\varphi + 3.443)/V_{0}$ \\
		$W_1$ & $(50.42\,\varphi - 16.74)/V_{0}$ \\
		$W_2$ & $(73.71\,\varphi - 32.20)/V_{0}$ \\
		\midrule
		$R_1$ & $(-0.1110\,\varphi + 0.12020)/R_{0}$ \\
		$R_2$ & $(-0.2042\,\varphi + 0.09266)/R_{0}$ \\
		$R_B$ & $(-0.05194\,\varphi + 0.2086)/R_{0}$ \\
		$R_C$ & $(-0.22420\,\varphi + 0.1125)/R_{0}$ \\
		\midrule
		$k_B$ & $(2.126\,\varphi + 1.238)/V_{0}^{2}$ \\
		$k_C$ & $(-2.164\times10^{4}\varphi^{3} + 2.378\times10^{4}\varphi^{2} - 8722\varphi + 1072)/V_{0}^{2}$\\
		\bottomrule
	\end{tabular}
	\caption{Curve-fit expressions for baseflow parameters as functions of the flow coefficient $\varphi \in [0.338,\ 0.408]$  where $R_0 = 0.2$ m, $V_{0} = 1000\times 2\pi/60 \times R_0$ and $\Omega_0 = V_0/R_0$ (from \cite{Susan10}).}
\end{table}

\end{document}